\documentclass[sigconf]{acmart}
\AtBeginDocument{%
  }

\setcopyright{acmlicensed}
\copyrightyear{2025}
\acmYear{2025}
\acmDOI{XXXXXXX.XXXXXXX}
\acmConference[ACM CCS 2025]{32nd ACM Conference on Computer and Communications Security (CCS)}{October 13-17, 2025}{Taipei, Taiwan}
\acmISBN{978-1-4503-XXXX-X/2018/06}

\usepackage{subfiles}
\usepackage{soul}
\usepackage{xcolor}
\usepackage{graphicx}
\usepackage{colortbl}
\usepackage{multirow}
\usepackage{framed}

\begin{document}

\title{Conflicting Scores, Confusing Signals: An Empirical Study of Vulnerability Scoring Systems} 






\DeclareRobustCommand{\okina}{%
  \raisebox{\dimexpr\fontcharht\font`A-\height}{%
    \scalebox{0.8}{`}%
  }%
}

\author{Viktoria Koscinski}
\affiliation{%
  \institution{Rochester Institute of Technology}
  \city{Rochester}
  \state{NY}
  \country{USA}}
\email{vk2635@rit.edu}

\author{Mark Nelson}
\affiliation{%
  \institution{University of Hawai\okina i at Mānoa}
  \city{Honolulu}
  \state{HI}
  \country{USA}}
\email{marknels@hawaii.edu}

\author{Ahmet Okutan}
\affiliation{%
  \institution{Leidos}
  \city{Reston}
  \state{VA}
  \country{USA}}
  \email{ahmet.okutan@leidos.com}

\author{Robert Falso}
\affiliation{%
 \institution{Rochester Institute of Technology}
  \city{Rochester}
  \state{NY}
  \country{USA}}
 \email{rf8580@rit.edu}

\author{Mehdi Mirakhorli}
\affiliation{%
  \institution{University of Hawai\okina i at Mānoa}
  \city{Honolulu}
  \state{HI}
  \country{USA}}
  \email{mehdi23@hawaii.edu}

\renewcommand{\shortauthors}{Koscinski et al.}

\begin{abstract}
 Accurately assessing software vulnerabilities is essential for effective prioritization and remediation. While various scoring systems exist to support this task, their differing goals, methodologies and outputs often lead to inconsistent prioritization decisions. 
 This work provides the first large-scale, outcome-linked empirical comparison of four publicly available vulnerability scoring systems: the Common Vulnerability Scoring System (CVSS), the Stakeholder-Specific Vulnerability Categorization (SSVC), the Exploit Prediction Scoring System (EPSS), and the Exploitability Index. We use a dataset of 600 real-world vulnerabilities derived from four months of Microsoft's Patch Tuesday disclosures to investigate the relationships between these scores, evaluate how they support vulnerability management task, how these scores categorize vulnerabilities across triage tiers, and assess their ability to capture the real-world exploitation risk. Our findings reveal significant disparities in how scoring systems rank the same vulnerabilities, with implications for organizations relying on these metrics to make data-driven, risk-based decisions. We provide insights into the alignment and divergence of these systems, highlighting the need for more transparent and consistent exploitability, risk, and severity assessments.
\end{abstract}





\received{14 April 2025}
\received[revised]{12 March 2009}
\received[accepted]{5 June 2009}

\maketitle

\section{Introduction}\label{sec:introduction}
Vulnerability Management (VM) is the process of identifying, classifying, remediating, and mitigating vulnerabilities \cite{foreman2019vulnerability}. It consists of several interrelated activities: 
\textit{discovery and research}, identifying previously undiscovered vulnerabilities; 
\textit{report intake}, receiving and processing information about vulnerabilities; 
\textit{analysis}, developing an understanding of a vulnerability's potential impact, root causes, and remediation/mitigation strategies; 
\textit{coordination}, sharing information among stakeholders and those involved in disclosure; 
\textit{disclosure} to constituents enabling informed decisions; and 
\textit{response}, including remediating, mitigating, or patching vulnerabilities \cite{kossakowski2019computer}.

\textit{Vulnerability prioritization} is fundamental for security practitioners \cite{ssvc} as organizations have limited resources while the gross number of vulnerabilities discovered grows monotonically. One of the most utilized VM resources is the US National Vulnerability Database (NVD) \cite{nvd}, which contains data about vulnerabilities that are assigned Common Vulnerabilities and Exposures (CVE) IDs. Recent NVD trends show that the number of published CVEs has grown significantly with 25,059, 28,961, and 29,004 CVEs published in 2022, 2023, and 2024, respectively \cite{cve_number}.
This overwhelming volume of vulnerabilities and alerts creates a significant operational challenge. One survey of over 600 cybersecurity professionals found that 63\% are unable to act on the large number of alerts, and 67\% feel they do not have the time to mitigate all vulnerabilities \cite{Ponemon2020VulnMgmt}. The study explicitly states the desire from respondents for ``a risk-based and prioritized list of actions,’’ directly advocating for the role of scoring systems.

To manage the scale of vulnerabilities, the state-of-the-practice relies on vulnerability scoring systems to measure their severities and drive actions for effective outcomes~\cite{hans2022best}.
Several vulnerability scoring systems have been created with the goal of providing insights for security practitioners. 
The most widely used is the Common Vulnerability Scoring System (CVSS) \cite{fruhwirth2009improving, spring2021time, cvss3.1, howland2023cvss}, with CVSS scores available in the NVD. 

While CVSS is currently the industry standard, security researchers question its use for vulnerability prioritization \cite{spring2021time, howland2023cvss, spring2018towards-improving}. Other scoring systems have emerged to address CVSS's gaps, these include CISA's emerging Stakeholder-Specific Vulnerability Categorization (SSVC)~\cite{ssvc}, the Exploitability Index~\cite{exp_index_2022}, and the Exploit Prediction Scoring System (EPSS)~\cite{epss,epss-paper} which is developed by FIRST organization. Many other scoring systems  are proprietary and not publicly available \cite{vmdr, nexpose, recorded-future, snyk, vpr1}, or have very specific use cases and only score a subset of vulnerabilities \cite{microsoft-exploitability, ivss, rss}. 

Although prominent scoring systems have been individually studied and critiqued \cite{croft2022investigation, jiang2021evaluating, milousi2024evaluating, jacobs2021exploit}, much of the existing research is qualitative, anecdotal, or narrowly focused, providing documentation-based reviews of scope and limitations \cite{spring2021time, howland2023cvss}. Other studies have examined theoretical underpinnings, pointing to a lack of justification for the CVSS formula~\cite{spring2018towards-improving, ssvc-guide-cisa}, vague specifications \cite{spring2021time}, and skewed score distributions \cite{howland2023cvss}. While some analyses have investigated inconsistencies, they have typically focused on inter-rater variability when scoring the same vulnerability with a single system like CVSS \cite{holm2015expert, croft2022investigation}, or have examined individual systems’ score distributions in isolation.

\textbf{Position of This Paper:}  Critically, prior work has not been tied to the operational effectiveness of these systems for real-world vulnerability management. No study has empirically compared the uniform messaging, predictive value, and practical utility of multiple, competing scoring systems—such as CVSS, EPSS, and SSVC—when applied to a shared set of real-world vulnerabilities from an industrially relevant context. While critiques have inspired alternatives like SSVC~\cite{ssvc, bulut2022vulnerability}, the field lacks a large-scale, data-driven comparison to determine whether newer systems are more effective or how their recommendations align or conflict with established ones. An empirical study is therefore necessary to provide measurable, grounded evidence of score disparities that prior qualitative studies cannot, offering specific insights beyond theory to reveal hidden patterns, challenge assumptions, and quantify these systems’ true impact on prioritization and remediation decisions~\cite{suciu2022expected}.

Therefore, in this paper we conduct a comprehensive \textbf{empirical study of Microsoft’s Patch Tuesday} disclosures to compare the effectiveness and practicality of four vulnerability scoring systems. Patch Tuesday represents a recurring and high-impact event in vulnerability management, making it a meaningful lens through which to evaluate how scoring systems inform prioritization and remediation decisions. We apply each scoring system to 600 real-world vulnerabilities disclosed by Microsoft over a four-month period and analyze their messaging, consistency, triage support, and exploitability signals. Our goal is to assess how these systems (CVSS, SSVC, EPSS, and Exploitability Index) differ in practice, and how effectively they guide security practitioners in prioritizing response and triage efforts. 

\noindent To this end, we investigate the following \textbf{research questions}:\vspace{-5pt}
\begin{itemize}
    \item \textbf{RQ1:} \textit{How consistent is the messaging across vulnerability scoring systems during Patch Tuesday?} We measured  inter-system agreements and found that the  four scoring systems exhibit very low correlation and agreement with each other, indicating that the messaging they provide is inconsistent. This means a CVE’s score and  its perceived severity or urgency can vary significantly depending on which system is used, complicating unified and meaningful triage decisions.

    \item \textbf{RQ2:} \textit{How do scoring systems differ in their ability to support triage and patch prioritization efforts?} 
    We found that scoring bins result in difficulty deciding which CVEs to prioritize due to the high number of CVEs in a limited set of bins. Furthermore, scoring systems do not agree on the top N CVEs, further supporting the previous empirically grounded findings about scoring inconsistencies but also resulting in implications for the use of more than one scoring system to support triage. 

    \item \textbf{RQ3:} \textit{How well do time-based exploit prediction scores (EPSS) align with actual exploitation events compared to static scoring systems?} We found that EPSS rarely predicted exploitability, contrary to the premise of the approach. The time-based analysis of known exploited vulnerabilities demonstrated that EPSS fails to predict or measure likelihood of exploitation with high confidence before CVEs addition to the CISA KEV catalog; fewer than 20\% of CVEs ever exceeded a 50\% chance of exploit beforehand. In contrast, static scoring systems, particularly CVSS, had a tendency to assign higher severity scores to CVEs later confirmed as exploited.

    \item \textbf{RQ4:} \textit{Do scoring systems behave differently for different vulnerability types?} CVEs are tagged with Common Weakness Enumeration (CWE) data \cite{cwe}, representing the underlying software weakness that contributes to each vulnerability. We found that the way scoring systems treat different vulnerability types shows no universal patterns in scoring agreement. In other words, scoring behavior is largely independent of CWE classification.
\end{itemize}

The \textbf{contributions} of our work are five-fold:

\begin{itemize}
\item
\textbf{Industry-Grounded Empirical Study} – We conduct the \textbf{first large-scale, empirical comparison} of four \textbf{vulnerability scoring systems widely used in practice} (CVSS, EPSS, SSVC, and the Exploitability Index) on a real-world, high-stakes dataset of 600 \textbf{Microsoft Patch Tuesday} CVEs. The dataset mirrors enterprise patch cycles.

\item We provide a comprehensive, data-driven empirical research framework that \textbf{examines consistency}, \textbf{triage effort}, \textbf{actionability}, and \textbf{exploit prediction} alignment.

\item \textbf{Novel Operational Metrics}: This paper introduces novel, operationally relevant metrics—\textit{bin-based triage effort} and \textit{top-N overlap}—to quantify analyst workload, prioritization agreement, and exploitation alignment, providing actionable insights directly applicable to real-world industrial vulnerability management.

\item The findings of the paper have \textbf{practical significance for the industry} and \textbf{inform academic communities} as they reveal significant divergence in scoring behavior, expose limitations in predictive systems like EPSS, and offer actionable insights for practitioners seeking to choose or combine scoring systems effectively.

\item An in-depth discussion of limitations, weaknesses and failure points of scoring systems requiring further investigation.
\end{itemize}

\noindent \textbf{Replication Package:} To support transparency and \textbf{reproducibility} of the findings, all collected data, evaluations, and source code
is available at: 
\url{https://github.com/SoftwareDesignLab/Vulnerability-Scoring-Systems-Comparison}. 


\section{Vulnerability Scoring Systems Studied}
\label{sec:background}
In this section, we provide a brief overview describing how each of our studied scoring systems rates vulnerabilities.

\subsection{Common Vulnerability Scoring System} \label{sec:cvss-background}
CVSS was developed to systematically characterize vulnerabilities and produce a numerical score that represents their severity. The score is based on a formula with discrete input parameters. It outputs a scalar score ranging from 0 (not vulnerable) to 10 (critical) in increments of 0.1 (resulting in 100 possible scores). Each CVSS score maps to a qualitative severity label—Low (0.1–3.9), Medium (4.0–6.9), High (7.0–8.9), or Critical (9.0–10.0)—to help organizations assess and prioritize vulnerabilities more effectively.

CVSS metrics assess the qualities of vulnerabilities. These metrics differ between CVSS v3 (currently most widely used) and v4 (the newest version). They consist of \textit{base metrics}, which are required for SSVC score generation and optional metrics.
Base metrics consist of \textit{exploitability} (attack vector, user interaction, and complexity), \textit{vulnerable system impact} in terms of confidentiality, integrity, and availability (CIA), and \textit{subsequent system impact} also in terms of CIA.
Environmental metrics override base metrics to express the importance of the affected IT asset to a user's organization.
Additional optional metrics include \textit{temporal metrics} (v3), \textit{threat metrics} (v4), and \textit{supplemental metrics} (v4) \cite{cvss3.1, cvss4}.  
Of the four scoring systems we study, CVSS is the most well-documented.
Additional information is provided on the CVSS website \cite{cvss4website}.

\begin{table}
    \centering
    \caption{Mission and Well-Being Impact \cite{ssvc-guide-cisa}.}
    \small 
    \begin{tabular}{c|p{1.4cm}|p{1.4cm}|p{1.4cm}|p{1.5cm}|}
         \multicolumn{2}{c}{}  & \multicolumn{3}{c}{\textit{Public Well-Being Impact}} \\ \cline{3-5}
         \multicolumn{2}{c|}{}  & \centering\textbf{minimal} & \centering\textbf{material} & \centering\textbf{irreversible} \tabularnewline
         \cline{2-5}
        \multirow{3}{*}{\rotatebox[origin=t]{90}{\parbox{1.2cm}{\centering\textit{Mission Prevalence}}}}  & \centering\textbf{minimal} & low & medium & high \\ \cline{2-5}
                                             & \centering\textbf{support} & medium & medium & high \\ \cline{2-5}
                                             & \centering\textbf{essential} & high & high & high \\ \cline{2-5}
    \end{tabular}
    \label{tab:m-and-wb}
\end{table}

\subsection{Stakeholder-Specific Vulnerability Categorization (SSVC)}
The Stakeholder-Specific Vulnerability Categorization (SSVC) is a decision tree model developed to improve vulnerability prioritization and mitigate perceived shortcomings of CVSS \cite{ssvc}.
The US Cybersecurity and Infrastructure Security Agency (CISA) developed a custom SSVC decision tree for vulnerability response for the US federal government, state/local governments, and critical infrastructure entities  \cite{ssvc-guide-cisa}. CISA's SSVC decision tree utilizes a qualitative evaluation of factors affecting a vulnerability's priority level, and outputs a \textit{priority label} indicating what action it recommends with respect to that vulnerability, in contrast to a numerical score like CVSS \cite{aranovich2021beyond, ssvc}. The goal is to help vulnerability managers decide what to do about a discovered vulnerability \cite{ssvc}.

Each decision point in SSVC's decision tree has at least two \textit{decision values}, which lead to a subsequent decision point, with the last decision point resulting in a final outcome, or \textit{priority label} about what action to take regarding the vulnerability. 
CISA's SSVC model was developed based on vulnerabilities relevant to various critical infrastructure entities \cite{cisa-website-ssvc}. 
This decision tree's four decision points are described below, based on CISA's SSVC guide \cite{ssvc-guide-cisa}:

\textbf{(State of) Exploitation} describes the vulnerability's present state of exploitation.
It has three possible decision values. \textit{None} indicates no evidence of active exploitation or public proof of concept. \textit{Public proof-of-concept (PoC)} indicates that there is either a publicly available PoC on the Web or the vulnerability has a well-known method of exploitation. \textit{Active} describes vulnerabilities with reliable evidence that they have been exploited by attackers in the wild.

\textbf{Technical Impact} describes the control gained over, or the information exposed about, the vulnerable component. It has two possible decision values: \textit{partial} when an adversary  obtains limited control over or information about the software with the vulnerability (if exploited), and 
\textit{total} when an adversary gains total control over the software or total information disclosure.  

\textbf{Automatable} describes whether a vulnerability's exploit may be automated, and has two possible decision values. \textit{No} refers to cases where the first four steps of the cyber kill chain~\cite{cyber-kill-chain}
cannot be automated for the vulnerability. \textit{Yes} refers to the case where these steps can be automated or where there are no known barriers to automation.

\textbf{Mission and Well-Being Impact} 
is a combined decision based on both the \textbf{mission prevalence} and the \textbf{public well-being impact}. Mission prevalence describes the effects of a vulnerability on mission-essential functions. It can be \textit{minimal}, \textit{support}, or \textit{essential}.
Public well-being impact describes the effects of a vulnerability on the affected system's operators or consumers as defined by the Centers for Disease Control and Prevention \cite{cdc}. It can be \textit{minimal}, \textit{material}, or \textit{irreversible}.
The resulting values are shown in Table \ref{tab:m-and-wb}.

CISA also defines four priority labels based on combinations of decision values. 
Table~\ref{tab:vuln_decision} provides an overview of the priority labels.

\begin{table*}[h!]
    \vspace{-10pt}
\centering
\caption{SSVC Priority Labels.}    \vspace{-10pt}
\label{tab:vuln_decision}
\scriptsize
\begin{tabular}{|>{\centering\arraybackslash}m{1.5cm}|>{\arraybackslash}m{15cm}|}
\hline
\textcolor{green}{\textbf{Track}} & The vulnerability does not require action at this time. The organization would continue to track the vulnerability and reassess if new information becomes available. CISA recommends remediating \textbf{Track} vulnerabilities \textit{within standard update timelines}. \\
\hline
\textcolor{yellow}{
\textbf{Track*}} & The vulnerability contains specific characteristics that may require closer monitoring for changes. CISA recommends remediating \textbf{Track*} vulnerabilities \textit{within standard update timelines}. \\
\hline
\textcolor{orange}{
\textbf{Attend}} & The vulnerability requires attention from the organization's internal supervisory-level individuals. Necessary actions may include requesting assistance or information about the vulnerability and may involve publishing a notification, either internally and/or externally, about the vulnerability. CISA recommends remediating \textbf{Attend} vulnerabilities \textit{sooner than standard update timelines}. \\
\hline
\textcolor{red}{
\textbf{Act}} & The vulnerability requires attention from the organization's internal supervisory-level and leadership-level individuals. Additional information or assistance should be requested, and a notification should be published internally or externally. Remediate as soon as possible. \\
\hline
\end{tabular}
\end{table*}

\subsection{Exploit Prediction Scoring System}
The Exploit Prediction Scoring System (EPSS), like CVSS, was developed by FIRST.Org, Inc. \cite{epss}. 
It is designed to estimate the likelihood (probability) that a software vulnerability will be exploited in the wild within the next 30 days. 
EPSS aims to address other scoring systems' limited ability to assess threat, although it does not account for any specific environmental controls or estimate the impact of the vulnerability being exploited \cite{epss2}. 
Although FIRST.Org, Inc. does not share the underlying data, model and/or source code of EPSS, the general machine learning techniques used, as well as results, are published in an academic paper \cite{epss-paper}. 
EPSS takes into account data such as the vendor, age of the vulnerability, keywords in the vulnerability description, CVSS metrics, mentions of the vulnerability online, publicly available exploit code, and more. 

The EPSS model produces a score between 0 and 1, representing the probability that a vulnerability will be exploited in the next 30 days. Unlike CVSS and SSVC, EPSS does not assign qualitative category labels to the various percentages. Scores change over time and are calculated daily. EPSS scores for any day are available for download from the EPSS website\footnote{\url{https://www.first.org/epss/data_stats}}. This website also provides information on top rated recent CVEs, CVEs with shifting EPSS scores, distributions of EPSS scores across vendors, and a comparison of EPSS scores with CVSS scores. 

\subsection{Exploitability Index}
Exploitability Index \cite{exp_index_2022,ExploitabilityIndex-github}  was introduced as a learning based approach which takes advantage of both Convolutional Neural Network (CNN)-based prediction and a data-driven common product enumeration (CPE)-based scoring model. The Exploitability Index aims to assess the likelihood that a vulnerability is exploited in the wild, using publicly available descriptions from the NVD. By encoding vulnerability descriptions into semantic representations using CNNs, the trained model learns patterns linked to the availability of historical exploits. 

The CNN model was trained using the CVE descriptions from the NVD and corresponding exploit data from various exploit databases. Experimental evaluations and case studies demonstrated that CNN models can predict the severity of vulnerabilities with high confidence. This exploitability scoring method was chosen in our study as it has outperformed the existing exploitability scores provided by the NVD, suggesting a more effective means of assessing the potential risk associated with software vulnerabilities. To compute the exploitability score, the authors developed a composite metric that combines CNN-based predictions with an empirically derived Product Hygiene Index based on the CPE. This index is based on how often a given product (identified via CPEs) has been associated with exploited vulnerabilities in the past. The final exploitability score is derived by weighting the CNN's output (indicating the likelihood that a vulnerability will be exploited) with the historical exploit frequency of the affected product. Like CVSS, the resulting Exploitability Index produces scores on a 0–10 scale, but offers a more nuanced and  adaptive alternative to static, rule-based systems such as CVSS.


\vspace{-2pt}
\section{Case Study Setup}
To address our research questions from Section~\ref{sec:introduction}, we conducted an embedded case study~\cite{Runeson2009} of Microsoft’s Patch Tuesday disclosures, following established guidelines for empirical research~\cite{Verner2009}. We use a single embedded case study design — one case (Patch Tuesday), with multiple units of analysis (vulnerabilities across Microsoft products). This setup allows for diverse scoring behavior to be analyzed across 600 real-world CVEs.

\subsection{Case Selection}
\textbf{Patch Tuesday} occurs on the second Tuesday of each month at about 10 a.m. Pacific Standard Time. The Microsoft Security Response Center investigates all reports of security vulnerabilities affecting Microsoft products and services, and provides this information as part of an ongoing effort to manage security risks and help keep systems protected. 
We selected Microsoft's Patch Tuesday as the basis for our embedded case study due to its unique position in the software security ecosystem. Patch Tuesday represents a consistent, high-impact, and well-structured vulnerability disclosure and remediation event that occurs monthly. It provides a controlled and repeatable environment in which hundreds of vulnerabilities across a wide range of Microsoft products are disclosed simultaneously, often accompanied by vendor-supplied severity ratings and exploitability indicators.

This setting is particularly well suited for comparing vulnerability scoring systems, as all Patch Tuesday vulnerabilities are:
(i) released under similar timing and disclosure conditions,
(ii) well-documented in Microsoft's Security Update Guide, with most Patch Tuesday CVEs having a detailed Q\&A describing their impacts and exploitability, and
(iii) relevant to enterprise vulnerability management teams who must prioritize responses quickly and at scale.
This selection strategy allows us to focus on real-world scoring behavior under practical triage constraints while holding contextual variables, such as disclosure policy, vendor communication, and patch availability, constant.

\subsection{Data Collection}

Our dataset includes 600 vulnerabilities disclosed across four Patch Tuesday events between April and July 2024. These span multiple Microsoft product families, including Windows, Office, Edge, and Azure. They exhibit a range of severity levels, CWE types, and exploitability characteristics. Each CVE serves as an embedded unit of analysis within the broader context of coordinated vulnerability disclosure by a single vendor.
For each CVE, we obtained data from the NVD, such as the CVE ID, vulnerability description, and CVSS score/vector, as well as data from Microsoft's Security Update Guide, which contained more data such as information on exploitability and a Q\&A.

\subsection{Scoring Vulnerabilities}
To ensure a comprehensive evaluation of each vulnerability scoring system, we assigned scores to each of the 600 Patch Tuesday vulnerabilities using each scoring framework: CVSS, Exploitability Index, EPSS, and SSVC. Since each system has different methodologies and data sources for scoring vulnerabilities, our approach to obtaining scores for each system also varied accordingly.

\subsubsection{CVSS Scores}
Each CVE released in Microsoft Patch Tuesday receives a CVSS score by the Microsoft security team. These base scores are also published on NVD. It is important to note that while CVSS temporal and environmental scores may be calculated, they are considered organization- and time-specific, and are therefore not provided by default in the NVD. Each organization can choose to calculate these if desired. 

\subsubsection{Exploitability Index scores}
We obtained Exploitability Index scores using a publicly available automated tool that incorporates a pre-trained CNN model and the exploit frequency of the affected product. The score is derived by combining the CNN's output, which indicates the likelihood that a vulnerability will be exploited, and the exploit frequency of the affected product \cite{exp_index_2022}. The Exploitability Index scoring method relies on vulnerability descriptions and Common Platform Enumeration (CPE) information, both of which we retrieved from the NVD to ensure consistency. 

\subsubsection{EPSS scores}
Unlike the other scoring systems studied, EPSS scores are generated and updated on a daily basis. While the model itself is trained on proprietary data and is not publicly available, both daily and historical scores are downloadable from the EPSS website. To maintain temporal consistency with Patch Tuesday vulnerability disclosures, we obtained EPSS scores from July 9th, 2024, which was the release date of last Patch Tuesday in our dataset. 
Although EPSS scores are updated daily, some vulnerabilities are not scored immediately. As a result, we were able to obtain EPSS scores for 458 CVEs, with 142 not yet scored by EPSS when the scores were obtained. 
For the studies requiring a score from each of the four systems, CVEs without EPSS scores were removed to ensure a complete dataset. All CVEs were kept for analysis in Sections 4.2.1, 4.2.2, and 4.3.2.

\subsubsection{SSVC scores}  To obtain SSVC scores, each vulnerability was systematically assessed by a security team with two years of experience scoring vulnerabilities according to the official guidelines of SSVC~\cite{ssvc-guide-cisa}. As described in Section \ref{sec:background}, SSVC relies on four key decision points: \textit{(State of) Exploitation}, \textit{Technical Impact}, \textit{Automatable}, and \textit{Mission \& Well-Being Impact (M\&WB)}. For \textit{(State of) Exploitation},  the team classified vulnerabilities as ``Active'' if they were listed in CISA’s Known Exploited Vulnerabilities (KEV) catalog, ``PoC'' if they were associated with a common weakness that enables consistent exploitation\footnote{\url{https://certcc.github.io/SSVC/reference/decision_points/exploitation/}} or if a PoC was publicly available, and ``None'' otherwise. \textit{Technical Impact} was determined using both NVD vulnerability descriptions and Microsoft's Patch Tuesday Q\&A documentation. The scoring team categorized impact as ``Total'' for vulnerabilities described as  having total control or information gained (with the help of key-phrases such as ``remote code execution'') and ``Partial'' for cases where total control or information is not gained. ``Automatable'' status (Yes/No) was also determined based on NVD descriptions and the Patch Tuesday Q\&A, particularly leveraging Q\&A sections detailing exploitation techniques and prerequisites. For example, if an attack required specific user interaction (e.g., clicking a malicious link), it was classified as not automatable.
Since M\&WB is inherently organization-specific, the security team did not attempt to assign a single definitive value. Instead, they computed three separate SSVC scores, corresponding to all possible M\&WB values: ``Low,'' ``Medium,'' and ``High.'' This allows for greater and more organization-specific flexibility in interpretation by different stakeholders.

\subsection{Data Analysis}

\textbf{RQ1 (Scoring System Consistency):}
To address this question, we evaluated the degree of agreement among scoring systems using three complementary approaches: 
(i) t-SNE visualizations~\cite{van2008visualizing} to illustrate clustering and divergence in scoring behavior across systems;
(ii) normalized score comparison to illustrate individual scoring differences among scoring systems;
(iii) rank- and value-based correlation metrics, including Kendall’s Tau, Spearman’s Rho, and Pearson correlation~\cite{puth2015effective}, and; 
(iv) categorical agreement measures such as Cohen’s Kappa~\cite{5584447} and Krippendorff’s Alpha~\cite{krippendorff2022content}.

\textbf{RQ2 (Support for Triage and Patch Prioritization):}  
To answer this question, we perform a bin-based effort estimation analysis to evaluate how effectively each scoring system supports triage and prioritization. Specifically, we calculate triage load and prioritization density by analyzing the distribution of CVEs across score bins, with a focus on the number of vulnerabilities concentrated in the top-ranked bins (e.g., top-1, top-2, and top-3 bins).
We also perform a top-N effort distribution estimation, where instead of using top-ranked bins, we use top N ranked vulnerabilities to analyze distributions at a more fine-grained level.




\textbf{RQ3 (EPSS Exploit Estimation Power):} We address this question by conducting a detailed temporal analysis of EPSS scores prior to exploitation. Specifically, we compare historical EPSS predictions with the timelines of known exploited vulnerabilities as documented in the KEV catalog, assessing how well EPSS identifies threats before public confirmation of exploitation.
We also analyze the distributions of known exploited vulnerabilities as scored by all four scoring systems to provide a comparison of how each system treats such vulnerabilities.

\textbf{RQ4 (Scoring System Behavior for Different Vulnerability Types):}
We answer this question by analyzing consistency among scoring systems for CVEs grouped by their associated CWE identifiers. We identify the most frequently occurring CWEs in our dataset and focus our analysis on the top five to ensure meaningful sample sizes. To evaluate whether scoring systems behave differently across these vulnerability types, we conduct a t-SNE visualization of CVEs tagged with the top five CWEs, using marker shapes to distinguish CWEs and color to reflect inter-system score agreement. Additionally, we compute agreement metrics (Cohen’s Kappa and Krippendorff’s Alpha) across scoring system pairs for each CWE to assess consistency at the vulnerability type level compared to that of all CVEs. This multi-perspective analysis allows us to determine whether CVEs of the same CWE exhibit consistent scoring behavior or agreement across systems.

\begin{figure}[htp]
    \centering
    \includegraphics[width=1\linewidth]{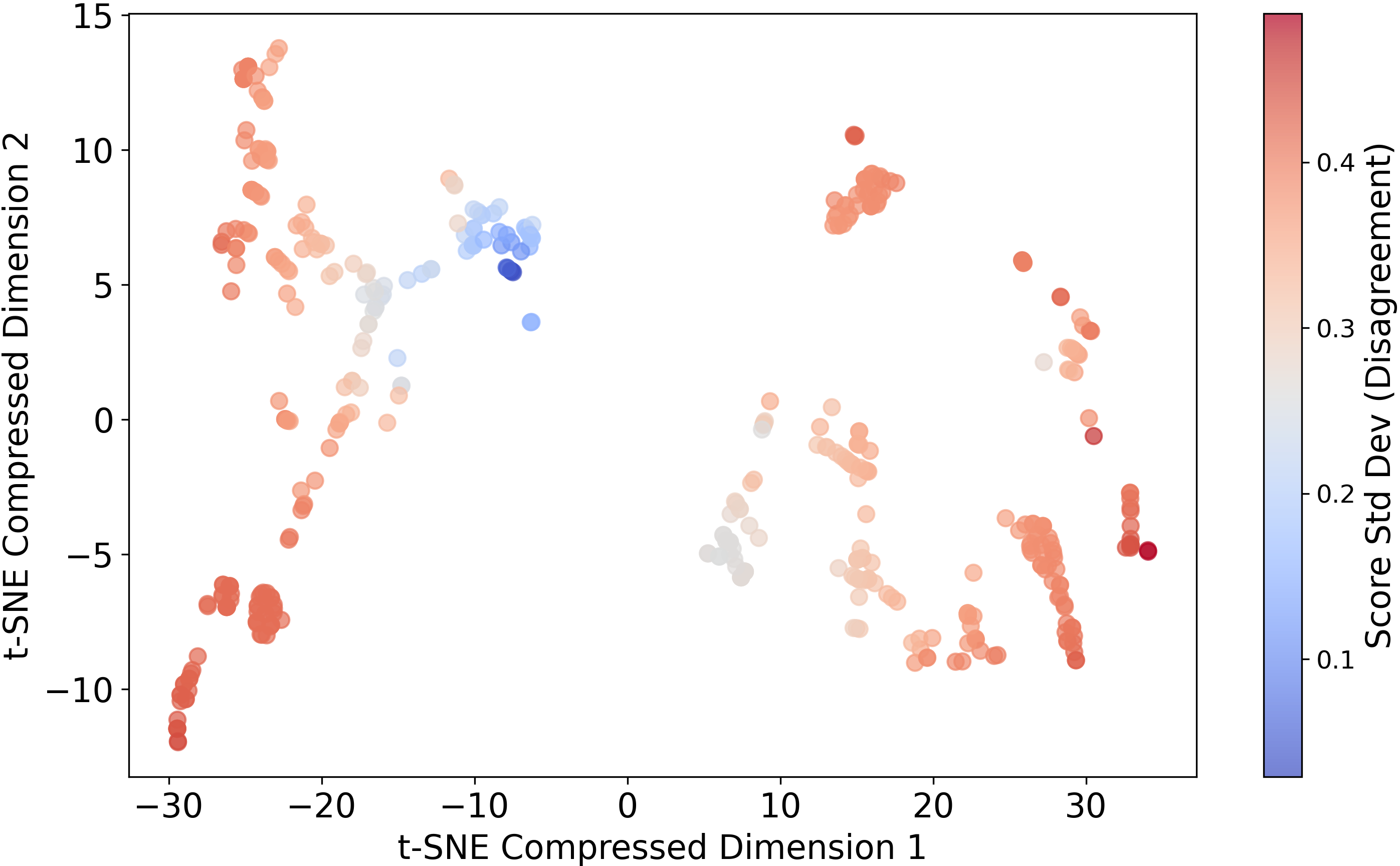}
    \caption{SSVC, CVSS, EPSS, and Exploitability Index scores visualized using t-SNE, colored by score agreement. The axes represent nonlinear dimensions computed by t-SNE to preserve relative similarity in the high-dimensional score space.}
    \label{fig:tsne}
\end{figure}
\section{Empirical Evaluation of Vulnerability Scoring Systems During Microsoft Patch Tuesday}\label{sec:evaluation}

In this detailed empirical study, we compare the messaging, effectiveness, and actionability of vulnerability scoring systems.

\subsection{RQ1: Evaluating Scoring System Messaging Consistency}

To evaluate the consistency of vulnerability scoring system messaging, we investigated how different scoring systems agree—or disagree—when rating the same vulnerabilities disclosed during Microsoft Patch Tuesday. Consistency is critical for coordinated triage and patching; inconsistent scores can lead to misaligned priorities across teams and tools.

\subsubsection{Visual Insight - t-Distributed Stochastic Neighbor Embedding (t-SNE) }
\begin{figure*}[!h]
    \centering

    \includegraphics[width=0.99\linewidth]{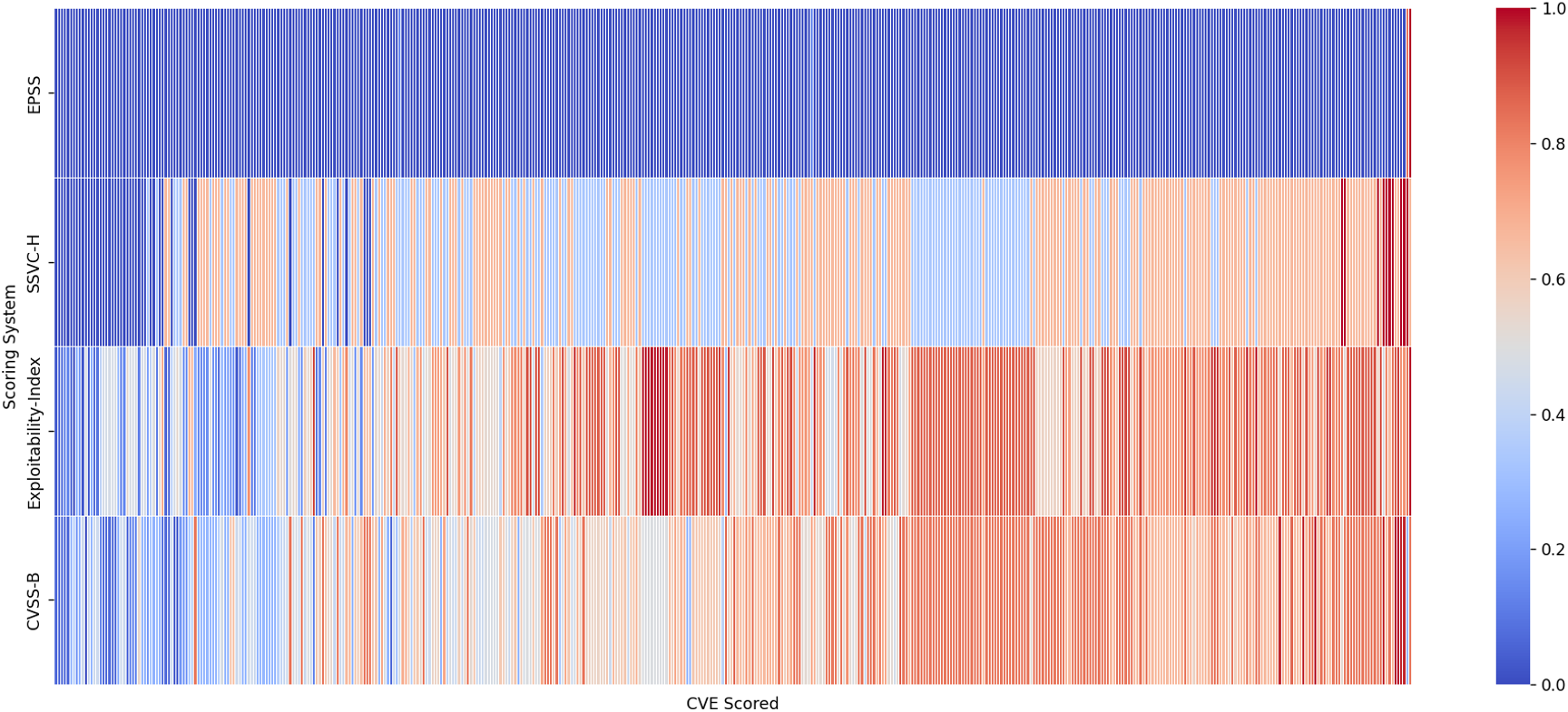}\vspace{-10pt}
    \caption{A heatmap of normalized scores for CVEs provided by each scoring system.}
    \label{fig:heatmap}
    \Description{A heatmap of normalized scores for CVEs provided by each scoring system.}
    
\end{figure*}
We begin by visualizing the scoring behavior of all four systems using t-Distributed Stochastic Neighbor Embedding (t-SNE) (Figure~\ref{fig:tsne}). 
t-SNE is a machine learning algorithm created for visualizing high-dimensional data using dimensionality reduction \cite{van2008visualizing}. Our t-SNE visualization takes all CVE points from six dimensions (CVSS base score, EPSS, SSVC-Low, SSVC-Medium, SSVC-High, and the Exploitability Index), and visualizes them in 2D in a way that preserves structure, colored by score agreement. Blue-colored areas represent CVEs with high agreement between scoring systems, while red-colored areas represent CVEs with strong disagreement between scoring systems. We removed 142 samples where EPSS did not provide a score. While EPSS scores CVEs much lower than any other scoring system, this does not significantly affect our t-SNE visualization results. 
\textbf{The tight blue cluster shows that only a small number of CVEs were scored similarly by all scoring systems.}\footnote{In order to examine if EPSS could skew the visualization data, we conducted the t-SNE visualization for a second time and removed the all EPSS data from the analysis, however the resulting visualization demonstrated the same divergence patterns. Here we only provide the t-SNE with all scoring systems.
} However, there are significantly more CVEs where the scoring systems diverge, as shown by the dark orange and red. These reddish tones (High Disagreement) indicate widespread disagreement across scoring systems. Additionally, the Wide Dispersion of CVEs and lack of a clear, dense clusters indicates that each scoring system contributes a unique signal, and there is no consistent pattern of agreement. We can conclude that, while some CVEs are reliably scored across systems, there are many regions in the data with systematic disagreement among the scoring systems analyzed.

\subsubsection{Normalized Score Comparison}
To complement the t-SNE view, we sort CVEs by their average normalized score across scoring systems and plot per-  system values (as shown in Figure \ref{fig:heatmap}). In this analysis, CVEs without EPSS scores were removed, resulting in 458 vulnerabilities in the figure, with each vertical line representing a unique CVE as scored by the four scoring systems. Next, the CVEs were sorted based on averages of normalized scores. As represented by the lines on the left and right edges of the figure, \textbf{only a handful of vulnerabilities had scores in agreement for all four systems.} Most CVEs in the middle portion of the figure show that scoring systems do not generally agree about how a given vulnerability should be scored.  It appears that CVSS and the Exploitability index have more agreement in scores with each other than with the other two scoring systems. However, there are still CVEs that are rated significantly higher in one scoring system than in the other. 

\vspace{-5pt}
\subsubsection{Correlation Coefficient Analysis}
To quantify these observations, we compute pairwise correlation coefficients between every two systems (Figure~\ref{fig:correlations}). We include Kendall’s Tau and Spearman’s Rho to capture ordinal agreement (ranking), and Pearson correlation for raw score comparison \cite{puth2015effective}. 
Kendall's Tau, which assesses whether there is a monotonic relationship between two variables by measuring ranking agreement, shows that CVSS has a week/moderate correlation with Exploitability Index and EPSS, while EPSS and Exploitability Index have a weak agreement. \textbf{All other combinations of unique scoring systems do not have a significant ranking agreement.} 

The stronger correlation between the SSVC scores provides some validation for this method as they are all related for low, medium, and high M\&WB. 
\textbf{Spearman's correlation coefficient shows similar trends}, with both metrics indicating that while there is some ordinal consistency, the scoring systems generally follow different ranking patterns. 
Pearson's correlation, which captures linear relationships, indicates a moderate association between CVSS-B and the Exploitability Index, \textbf{while its correlation with other scoring systems is near zero}, suggesting no linear relationship. 
Overall, the weak correlations across most pairs suggest that different scoring systems prioritize distinct aspects of vulnerability assessment. 

We obtained the p-values for our correlation analysis and found that for both Kendall's Tau and Spearman’s Rho, p-values are $<.05$ for all combinations of scores except for the Exploitability Index with SSVC-Low and SSVC-Medium. In contrast, the Pearson correlation yielded higher p-values for approximately half of the score combinations, especially those involving the Exploitability Index. This suggests a lack of linear relationships between some scoring systems. These findings also support the notion that Kendall's Tau and Spearman’s Rho, which measure monotonic relationships and rank-order agreement, align more closely with how prioritization scores are interpreted.


\subsubsection{Agreement Metrics}

We further analyze inter-system agreement using Cohen’s Kappa and Krippendorff’s Alpha after binning each system’s scores into categorical levels. This analysis supplements our prior correlation analysis, as both Cohen's Kappa and Krippendorff's Alpha quantify the extent of agreement beyond chance of two given scoring systems. 
Since CVSS, the Exploitability Index, and EPSS provide decimal number scores, and SSVC provides four categorical scores, we calculated these metrics based on scoring bins.  Comparisons with SSVC were binned using a four-quarter split.  Non-SSVC comparisons were placed into ten equally sized bins for a finer-grained comparison. The results, as shown in Table \ref{tab:alpha_kappa}, indicate a low level of agreement, if any, between the various scoring systems. \textbf{Cohen's Kappa values are close to zero across all comparisons, suggesting no agreement beyond chance. Similarly, Krippendorff's Alpha values are predominantly negative, indicating inconsistencies in how scoring systems classify vulnerabilities}. Even within the same scoring system, SSVC scores with different M\&WB values still have relatively weak agreement. These findings suggest that the scoring systems assess vulnerabilities from fundamentally different perspectives, highlighting a need for careful interpretation when selecting a scoring methodology to integrate. 

\begin{table}
    \centering
    \caption{Cohen's Kappa and Krippendorff's Alpha for each combination of scoring systems, including SSVC with low, medium, and high M\&WB.}\vspace{-10pt}
    \small
    \begin{tabular}{|c|c|c|}
    \hline
    \scriptsize
    \textbf{Scoring Systems} & \textbf{Kappa} & \textbf{Alpha} \\\hline
    CVSS - Exploitability Index & 0.01 & -0.03 \\\hline
    CVSS - EPSS  & 0.00 & -0.45 \\\hline
    CVSS - SSVC Low & 0.00 & -0.59 \\\hline
    CVSS - SSVC Medium & 0.00 & -0.52 \\\hline
    CVSS - SSVC High & -0.01 & -0.21 \\\hline
    Exploitability Index - EPSS & 0.00 & -0.41 \\\hline
    Exploitability Index - SSVC Low & 0.01 & -0.42 \\\hline
    Exploitability Index - SSVC Medium & -0.01 & -0.42 \\\hline
    Exploitability Index - SSVC High & 0.03 & -0.03 \\\hline
    EPSS - SSVC Low & 0.00 & -0.01 \\\hline
    EPSS - SSVC Medium & 0.02 & 0.00 \\\hline
    EPSS - SSVC High & 0.00 & -0.49 \\\hline
    \end{tabular}
    \normalsize
    \vspace{-10pt}
    \label{tab:alpha_kappa}
\end{table}


\begin{framed}
\vspace{-5pt}
    \noindent\textbf{Key findings}: 
    \begin{itemize}
    \item     We found that scoring systems exhibit low correlation and minimal agreement when assessing the same vulnerabilities. 
    \item Visualizations (e.g., t-SNE) show widespread divergence, and statistical measures—including Spearman's Rho, Kendall's Tau, Cohen’s Kappa, and Krippendorff’s Alpha—consistently confirm that no pair of systems aligns reliably. 
    \item This inconsistency suggests that the perceived severity and prioritization of a CVE may vary significantly depending on the chosen scoring system, complicating triage and decision-making processes.
 \end{itemize}
\end{framed}

\subsection{RQ2: Evaluating Scoring System Prioritization and Triage Using Effort Estimation}

To evaluate how well each scoring system supports patch prioritization and triage, we use effort estimation as a proxy for operational burden. Our goal is to assess whether a scoring system offers meaningful prioritization—guiding security teams toward high-impact vulnerabilities first without overloading them with noise.

We estimate triage effort by measuring how many vulnerabilities fall into the highest-priority bins defined by each system. However, effectiveness is not determined solely by placing fewer CVEs in the top bin. Rather, we assess whether the system provides a \textit{useful prioritization gradient} that allows security teams to progressively allocate attention and resources across bins of decreasing urgency.

\subsubsection{Bin-Based Effort Estimation}
We assess triage effort using a bin-based analysis, where each scoring system's outputs are divided into ordered priority bins. For CVSS, EPSS, and the Exploitability Index, we define 10 evenly spaced bins based on their respective score ranges (see Table~\ref{tab:bins-cvss-ei-epss}), allowing for a consistent comparison across systems. For SSVC, which provides four discrete decision categories—Track, Track*, Attend, and Act—we treat the ``Act'' and ``Attend'' outcomes as the top priority bins (see Table~\ref{tab:vuln_decision}).

To estimate effort, we measure how many vulnerabilities each system places in its top one, two, or three bins. This reflects the volume of CVEs a practitioner would need to address if following the system’s highest-priority recommendations. However, raw counts alone do not indicate effectiveness: a scoring system that places only a few CVEs in the top bin but many more in the second or third may not provide actionable stratification. Therefore, our analysis also considers whether each system produces a ``meaningful prioritization gradient''—that is, whether its binning structure helps practitioners phase their response across tiers of urgency, rather than forcing binary all-or-nothing decisions. This approach highlights the extent to which scoring systems support progressive, scalable triage under realistic operational constraints.



\begin{table}
    \centering
    \caption{Scoring bin ranges considered for CVSS, Exploitability Index, and EPSS, where $x$ is the normalized CVE score.}\vspace{-10pt}
    \small
    \begin{tabular}{|c|c|c|}
    \hline
        \textbf{CVSS} & \textbf{Exploitability Index} & \textbf{EPSS} \\\hline
        \multicolumn{2}{|c|}{$9\leq x\leq10$}  & $0.9\leq x\leq1.0$ \\\hline
        \multicolumn{2}{|c|}{$8\leq x<9$}  & $0.8\leq x<0.9$ \\\hline
        \multicolumn{2}{|c|}{$7\leq x<8$}  & $0.7\leq x<0.8$ \\\hline
        \multicolumn{2}{|c|}{$6\leq x<7$}  & $0.6\leq x<0.7$ \\\hline
        \multicolumn{2}{|c|}{$5\leq x<6$}  & $0.5\leq x<0.6$ \\\hline
        \multicolumn{2}{|c|}{$4\leq x<5$}  & $0.4\leq x<0.5$ \\\hline
        \multicolumn{2}{|c|}{$3\leq x<4$}  & $0.3\leq x<0.4$ \\\hline
        \multicolumn{2}{|c|}{$2\leq x<3$}  & $0.2\leq x<0.3$ \\\hline
        \multicolumn{2}{|c|}{$1\leq x<2$}  & $0.1\leq x<0.2$ \\\hline
        \multicolumn{2}{|c|}{$0\leq x<1$}  & $0.0\leq x<0.1$ \\\hline
    \end{tabular}
    \vspace{-10pt}
    \label{tab:bins-cvss-ei-epss}
\end{table}

\begin{figure}
    \centering
    \includegraphics[width=\linewidth]{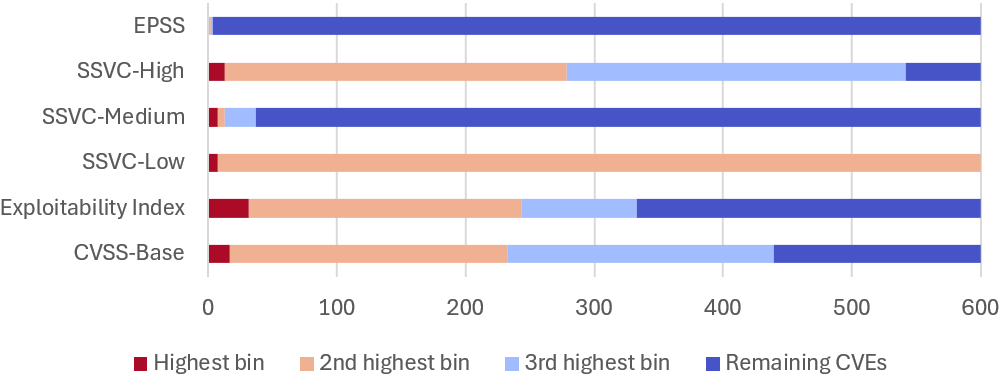}
    \caption{Prioritized CVEs in the top three bins for each scoring system.}
    \label{fig:bin-categorization}
    \Description{A chart showing the number of prioritized CVEs according to the top three bins in each scoring system.}
        \vspace{-10pt}
\end{figure}

In our analysis we ignore bins with zero vulnerabilities. For example, EPSS has one CVE in the 0.9-1.0 range, no CVEs in the 0.8-0.9 range, and one CVE in the 0.7-0.8 range. In this case, the highest bin is  0.9-1.0 and the second highest bin is 0.7-0.8. 
Figure \ref{fig:bin-categorization} triages 600 CVEs, binning them to prioritize which CVEs should be patched first. 
We assume that organizations will patch vulnerabilities in the first bin, then the second bin and finally the third bin, etc.

EPSS categorization results in only four CVEs being patched based on the top three bins. SSVC-High, SSVC-Low, Exploitability Index, and CVSS result in >50\% of the CVEs needing to be patched, effectively overwhelming security teams and undermining the purpose of triage by failing to concentrate attention on the highest-risk subset. SSVC-Medium provides the most reasonable effort estimate, with 8, 13, and 37 CVEs in the top 3 bins. 
Binning strategies we considered but did not analyze include (1) splitting the scoring bins into four quarters to match SSVC and (2) aligning CVSS binning with the severity labels described in Section \ref{sec:cvss-background}. These strategies produced large, coarse grained bins and did not allow for as much nuance in comparing the scoring systems. This is because larger bins retain less of the original scores' variabilities, thus making it more difficult to compare scoring systems and analyze any patterns that may emerge.


\subsubsection{Top-N Overlap}

We measured the overlap between the top N scored vulnerabilities. 
This analysis highlights two qualities between scoring systems: agreement in terms of prioritization, and the number of ``tied'' scores.
To conduct this analysis, we:
\begin{enumerate}
    \item Generate rankings for all CVEs in each scoring system. 
    \item Order CVEs for each scoring system by rank.
    \begin{itemize}
        \item Tied scores result in a tied rank. For example, CVSS has six 9.8 CVEs, so they all have a rank of 1.
    \end{itemize}
    \item Obtain the top $N$ CVEs for each scoring system, adding tied CVEs if needed. For example, if we obtain the top CVE ($N=1$) for CVSS, we include all six CVEs with a rank of 1. 
    \item Compare sets of top $N$ CVEs between scoring systems, counting the number of overlapping CVEs (the union of the sets). 
\end{enumerate}

As shown in Figure~\ref{fig:top-n}, agreement among scoring systems on top-ranked vulnerabilities is minimal: across the top 10 to top 100 CVEs, only five vulnerabilities are shared by all four systems. Pairwise comparisons reveal similarly weak overlap—for example, the Exploitability Index shares fewer than 20 CVEs with EPSS and SSVC-High within their respective top 100 lists.

Although CVSS appears to overlap more with EPSS and SSVC-High, this is largely due to extensive score ties, rather than true agreement. Specifically, CVSS assigns the same score to 198 CVEs in its top 20, meaning that 178 CVEs share the same score as the 20th-ranked one. Similar tie patterns exist in EPSS and SSVC-High, which include 102 and 279 CVEs, respectively, in their top 50 and top 20 score tiers. These large tie groups result in broad, undifferentiated priority bins, offering little actionable guidance for analysts seeking to triage vulnerabilities with precision.

\begin{framed}
\vspace{-5pt}
    \noindent\textbf{Key findings}: 
    \begin{itemize}
    \item  Scoring systems show minimal agreement on which CVEs to prioritize: only 5 CVEs overlap across all four systems in their top-100 lists. Large tie groups—such as 198 CVEs sharing a top-20 score in CVSS—further undermine the ability to rank vulnerabilities meaningfully. 
    \item Both bin-based and top-N analysis demonstrate that these scoring systems provide limited triage guidance, leaving security teams with broad, undifferentiated lists and no clear path for action. 
 \end{itemize}
\end{framed}

While using top-N analysis makes it easy to visualize how many CVEs are most important according to each scoring system, it is also apparent that if top-N analysis is used for CVE patch prioritization, the issue of having many CVEs with identical scores still exists. 
This leaves analysts with no clear guidance on how to prioritize such CVEs. 
Furthermore, the lack of agreement between scoring systems means that if multiple scoring systems are used, CVE prioritization is still unproven. 

\begin{framed}
 \noindent\textbf{Key findings}: 
    \begin{itemize}
    \item 
Our findings highlight trade-offs between different scoring systems when used for vulnerability triage. Scoring systems such as EPSS highlight few high-priority CVEs, and may overlook many CVEs which do not have the highest score. Systems such as SSVC and CVSS, and the Exploitability Index flag a broad set of CVEs as a priority, potentially overburdening response efforts. 
\end{itemize}
\end{framed}

\subsection{RQ3: Evaluating Scoring System Alignment with Real-World Exploitation}

\subsubsection{Distributions of Exploited Vulnerabilities.}
In this section, we analyze the scoring systems in terms of exploitation states and predictions. First, we normalze the distributions of scores to a value between 0 and 1. 
Of the 600 Patch Tuesday CVEs, 13 are in the KEV and therefore known to be exploited. 
Table \ref{tab:KEV-dist} shows how each scoring system evaluated the 13 exploited CVEs. 

CVSS tends to rate exploited CVEs high, with 12 greater than 0.7, but because these CVEs are being actively exploited, we argue that the CVSS score should be > 9 for all 13. 
The Exploitability Index rates them lower still, and according to the definition of the Exploitability Index, all 13 should be rated at 1. 
For SSVC, we found that the distribution of exploited vulnerabilities is heavily influenced by the M\&WB parameter and not the fact that the CVE is being exploited. 
Finally, we found that EPSS rates exploited vulnerabilities overwhelmingly low. Nine had a score lower than 0.1, three were unscored, and only one had a score > 0.1. 


\begin{table}[ht]
    \centering
    \caption{Distribution of (normalized) scores for known exploited vulnerabilities across different scoring systems.}\vspace{-5pt}
    \scriptsize
    \definecolor{low}{RGB}{255,235,238}
    \definecolor{medlow}{RGB}{255,205,210}
    \definecolor{med}{RGB}{239,154,154}
    \definecolor{medhigh}{RGB}{229,57,53}
    \definecolor{high}{RGB}{198,40,40}
    \newcommand{\heatcell}[1]{%
        \ifnum#1=0 \cellcolor{white}{#1}%
        \else\ifnum#1<2 \cellcolor{low}{#1}%
        \else\ifnum#1<5 \cellcolor{medlow}{#1}%
        \else\ifnum#1<9 \cellcolor{med}{#1}%
        \else\ifnum#1<12 \cellcolor{medhigh}{#1}%
        \else \cellcolor{high}{#1}%
        \fi\fi\fi\fi\fi
    }

    \begin{tabular}{|c|c|c|c|c|c|c|}
        \hline
        \textbf{Score range} & \textbf{CVSS} & \textbf{Expl. Index} & \textbf{SSVC-L} & \textbf{SSVC-M} & \textbf{SSVC-H} & \textbf{EPSS} \\\hline
        $0.9\leq x\leq1.0$ & \heatcell{3} & \heatcell{0} & \heatcell{0} & \heatcell{8} & \heatcell{13} & \heatcell{0} \\\hline
        $0.8\leq x<0.9$    & \heatcell{3} & \heatcell{0} & \heatcell{0} & \heatcell{0} & \heatcell{0}  & \heatcell{0} \\\hline
        $0.7\leq x<0.8$    & \heatcell{6} & \heatcell{6} & \heatcell{0} & \heatcell{0} & \heatcell{0}  & \heatcell{1} \\\hline
        $0.6\leq x<0.7$    & \heatcell{0} & \heatcell{2} & \heatcell{8} & \heatcell{5} & \heatcell{0}  & \heatcell{0} \\\hline
        $0.5\leq x<0.6$    & \heatcell{1} & \heatcell{3} & \heatcell{0} & \heatcell{0} & \heatcell{0}  & \heatcell{0} \\\hline
        $0.4\leq x<0.5$    & \heatcell{0} & \heatcell{2} & \heatcell{0} & \heatcell{0} & \heatcell{0}  & \heatcell{0} \\\hline
        $0.3\leq x<0.4$    & \heatcell{0} & \heatcell{0} & \heatcell{0} & \heatcell{0} & \heatcell{0}  & \heatcell{0} \\\hline
        $0.2\leq x<0.3$    & \heatcell{0} & \heatcell{0} & \heatcell{0} & \heatcell{0} & \heatcell{0}  & \heatcell{0} \\\hline
        $0.1\leq x<0.2$    & \heatcell{0} & \heatcell{0} & \heatcell{0} & \heatcell{0} & \heatcell{0}  & \heatcell{0} \\\hline
        $0.0\leq x<0.1$    & \heatcell{0} & \heatcell{0} & \heatcell{5} & \heatcell{0} & \heatcell{0}  & \heatcell{9} \\\hline
    \end{tabular}
    \label{tab:KEV-dist}
\end{table}

\subsubsection{EPSS Scoring of Exploited Vulnerabilities.}

The lack of EPSS prediction further motivated us to study the EPSS independently.
We investigated whether EPSS scoring can effectively predict whether or not a vulnerability will be exploited. To conduct this study, we analyzed all 1226 CVEs in the KEV catalog at the time of the study. For each CVE, we obtained the EPSS score at the beginning of each month from May 2021 (the first available score) until December 2024.  Each CVE had up to 43 months of EPSS scores.
As we are interested in EPSS scores \textit{prior to the CVE being added to the KEV catalog}, we removed EPSS scores generated after the CVE was added to the KEV catalog. 

\begin{table}[h!]
    \centering
    \caption{Number and percent of CVEs in the KEV catalog with corresponding ranges of highest EPSS scores obtained.}
    \vspace{-3pt}\small
    \begin{tabular}{|c|c|c|}
    \hline
        \textbf{Highest EPSS Score} & \textbf{\# of CVEs} & \textbf{\% of CVEs}\\\hline
        $\geq$ 0.5 & 244 & 19.9\%\\\hline
        $\geq$ 0.6 & 211 & 17.2\%\\\hline
        $\geq$ 0.7 & 192 & 15.7\%\\\hline
        $\geq$ 0.8 & 156 & 12.7\%\\\hline
        $\geq$ 0.9 & 102 & 8.3\%\\\hline
    \end{tabular}
    \normalsize
    \label{tab:EPSS-KEV}\vspace{-3pt}
\end{table}

As shown in Table \ref{tab:EPSS-KEV}, less than 20\% of exploited CVEs were  \textit{ever} rated > 0.5 (50\% chance of exploitation in the next month) by EPSS at any time prior to being added to the KEV catalog.
Only 8.3\% of CVEs had an EPSS score > 0.9 at any time prior to appearing in the KEV catalog. This, along with the fact that 275 (22.4\%) of CVEs did not have any EPSS score prior to known exploitation, suggests that \textbf{EPSS may not be a reliable predictor of whether a vulnerability will be exploited}. 


\begin{framed}
\noindent\textbf{Key findings}: 
    \begin{itemize}
    \item 
EPSS underperforms as a predictive tool for real-world exploitation. While it is designed to estimate the likelihood that a CVE will be exploited in the next 30 days, our empirical analysis shows that it often fails to flag exploited vulnerabilities in advance:
\begin{itemize}
    \item Only 19.9\% of known exploited CVEs in the KEV catalog had an EPSS score > 0.5 before exploitation.

\item Just 8.3\% ever reached an EPSS score > 0.9.

\item Over 22\% of exploited CVEs had no EPSS score at all prior to exploitation.
\end{itemize}
\item In contrast, static scoring systems (e.g., CVSS, Exploitability Index) more consistently assigned high scores to exploited CVEs, despite lacking predictive modeling.

\item These findings suggest that while EPSS provides a probabilistic and dynamic signal, its current performance limits its reliability as a standalone predictor for exploitation-based prioritization.
 \end{itemize}
\end{framed}

\subsection{RQ4: Evaluating Scoring System Behavior for Different Vulnerability Types}

To explore whether different vulnerability scoring systems behave differently across vulnerability types, we analyzed CWE tags associated with 600 CVEs in our dataset. These CVEs were mapped to 91 unique CWEs. However, only seven CWEs were associated with more than 20 CVEs each, while the majority were mapped to only one or a few CVEs. This limits the conclusions that can be drawn from this data, as there are not enough samples for most CWEs to draw meaningful generalizations. 14 CVEs were either not mapped to any CWE or were tagged with ``NVD-CWE-noinfo,'' limiting the ability to draw broad conclusions for those cases as well.

\subsubsection{Top-5 CWE t-SNE Visualization.}

\begin{figure}
    \centering
    \includegraphics[width=1\linewidth]{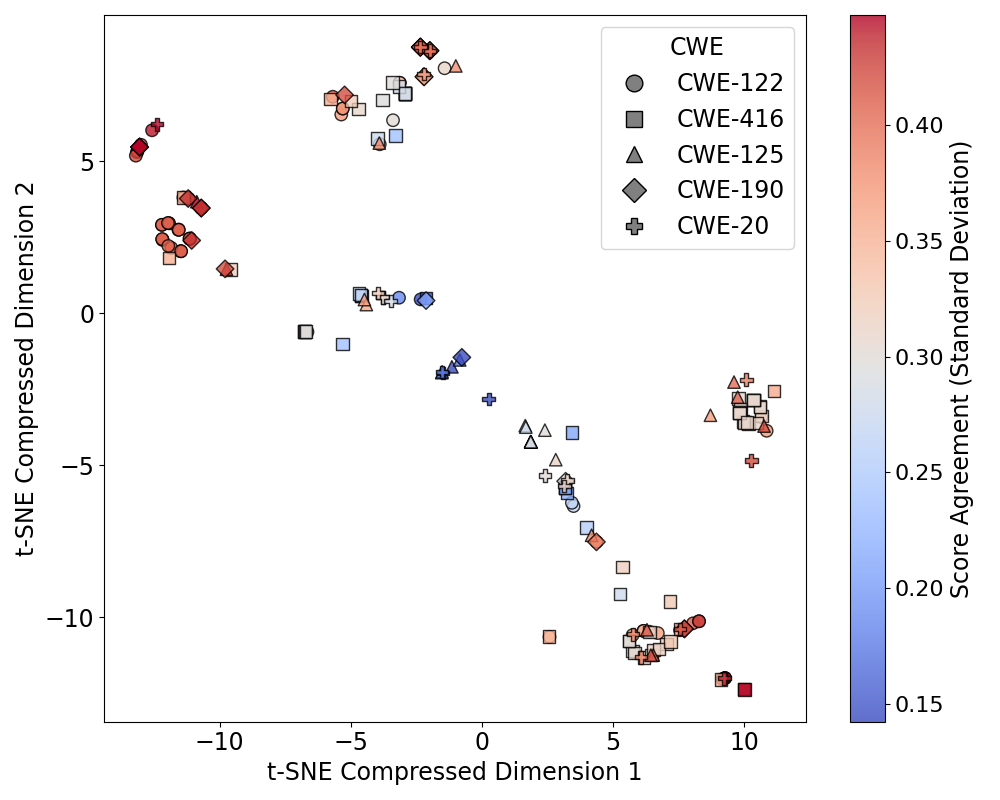}\vspace{-10pt}
    \caption{Scores for CVEs associated with the top five CWEs visualized using t-SNE.}
    \label{fig:CWE-tsne}
\end{figure}

Similar to before, we visualized the scoring behavior of all four systems using t-SNE (see Figure \ref{fig:CWE-tsne}). However, this time we focused on CVEs that are associated with the top five CWEs: \textit{CWE-122: Heap-based Buffer Overflow}, \textit{CWE-416: Use After Free}, \textit{CWE-125: Out-of-bounds Read}, \textit{CWE-190: Integer Overflow or Wraparound}, and \textit{CWE-20: Improper Input Validation}, containing 103, 81, 38, 31, and 27 samples, respectively. Coloring represents scoring agreement, while the shape of each point represents which CWE the point is mapped to. 
The distribution of shapes representing each CWE on the map shows that each of the top five CWEs has a wide range of scores across all systems, with different agreements. This suggests that scoring is not closely related to vulnerability types (CWEs) of the CVEs.
The small dark orange cluster on the upper left side of the map shows that there are some CVEs associated with CWE-122 that have a strong scoring agreement. However, since the points representing CVEs associated with CWE-122 are spread across the map, the scores and scoring agreements for these CVEs are still diverse, and being associated with CWE-122 will not  guarantee that score agreement is consistent. 
We can conclude that based on this visualization, the systematic disagreement among the scoring systems analyzed still exists among different vulnerability types, as exemplified by the wide variability, even though 
some CWEs demonstrated internal consistency in scoring across systems.
This suggests that no universal pattern exists in how scoring systems treat specific types of weaknesses.

\subsubsection{Agreement Metrics for CWEs.}
We further evaluated agreement between scoring systems at the CWE level, using Cohen's Kappa and Krippendorf's Alpha. While some CWEs showed greater agreement or disagreement than others, overall agreement trends did not significantly differ from the general population. In fact, the overall average of all unique scoring system pair agreements for all CWEs is consistent with the average of all unique scoring system pair agreements for the top five CWEs. 
These findings imply that, while vulnerability type (as captured by CWE) may influence scoring outcomes in isolated cases, it does not systematically affect the behavior of scoring systems across the dataset. This highlights the complexity of score interpretation and suggests that CWE context alone is insufficient for predicting how a vulnerability will be prioritized across different scoring paradigms.

\begin{table}
    \centering
    \vspace{-3pt}
    \caption{Average agreement score for Cohen's Kappa and Krippendorf's Alpha among top CWEs.}\vspace{-10pt}
    \small
    \begin{tabular}{|c|c|c|}
        \hline
        \textbf{CWE ID} & \textbf{Cohen's Kappa} & \textbf{Krippendorf's Alpha} \\\hline
        Average (All) & 0.01 & -0.29 \\\hline
        CWE-122 & 0.05 & -0.44 \\\hline
        CWE-416 & 0.07 & -0.33 \\\hline
        CWE-125 & 0.02 & -0.06 \\\hline
        CWE-190 & -0.06 & -0.41 \\\hline
        CWE-20 & 0.02 & -0.20 \\\hline
    \end{tabular}\normalsize\vspace{-10pt}
    \label{tab:my_label}
\end{table}


\begin{framed}
\noindent\textbf{Key findings}: 
    \begin{itemize}
\item We found that scoring systems do not behave consistently across specific vulnerability types. Our analysis of the top five CWEs most commonly associated with our CVE data revealed wide variability in scoring and score agreement for CVEs within the same CWE. While a small number of CVEs, particularly in CWE-122, showed higher scoring agreement, the majority were distributed widely, suggesting no systematic pattern in how scoring systems interpret vulnerability types. Agreement metrics further support this, with the average agreement for top CWEs closely matching the overall population. These results indicate that CWE context alone is insufficient to predict consistent scoring or prioritization outcomes across systems.
\end{itemize}
\end{framed}

\vspace{-7pt}
\section{Threats to Validity}

We follow established guidelines for evaluating empirical research in software engineering and discuss threats across three main dimensions: construct validity, internal validity, and external validity.

\textbf{Construct validity} concerns whether the evaluation metrics we use appropriately capture the concepts we intend to study, such as consistency, prioritization support, or exploit prediction quality.
Our study measures scoring consistency using correlation and agreement statistics, prioritization support through score diversity and triage bin analysis, and predictive performance via overlap with CISA’s KEV catalog. While these are standard and appropriate proxies, they do not capture all aspects of real-world decision-making. For example, KEV inclusion is a useful proxy for exploitation but does not reflect exploitability in all enterprise environments nor does it include 0-day vulnerabilities. Similarly, using score bins or top-N analysis to infer triage effort assumes organizations follow uniform workflows, which may not hold.
To mitigate these threats, we selected widely adopted metrics in the literature and carefully aligned each with the intended use case of the scoring system. 

\textbf{Internal validity} addresses whether our study’s conclusions are supported by sound data and analysis.
We ensured internal consistency by collecting scores from authoritative sources at a fixed time (April–July 2024), and by applying all scoring systems to the same set of 600 CVEs from Microsoft Patch Tuesday disclosures.
For our in-depth analysis of EPSS, we used all CVEs contained in the KEV catalog, which were frozen prior to KEV inclusion to avoid post-hoc bias in prediction evaluation. 
Nevertheless, we acknowledge two limitations: 
(i) EPSS scores can fluctuate daily; we captured snapshots rather than full time series. (ii) CVE mappings to KEV are binary (exploited or not) and may miss nuances such as partial exploit development or internal proof-of-concept exploits not in the wild.

\textbf{External validity} refers to the generalizability of our findings beyond the studied dataset.
Our dataset focuses exclusively on Microsoft vulnerabilities disclosed during Patch Tuesday over a four-month period. While this scenario represents a realistic, high-stakes environment where vulnerability prioritization occurs at scale, it may not reflect scoring behavior for non-Microsoft software, non-enterprise contexts, and CVEs with non-disclosure embargoes or vendor-specific triage paths.
Furthermore, we evaluated publicly available scoring systems only, and did not assess proprietary or context-aware tools used in some enterprise settings.
To mitigate this threat, we selected general-purpose scoring systems 
with broad adoption across public and private sectors, and we analyzed real-world CVEs from one of the most structured vulnerability disclosure streams in industry.

\section{Discussion}
A central finding of our study is the profound lack of agreement among prominent scoring systems. In this section we provide a detailed discussion and key insights on the root causes of such inconsistencies, and provide actionable recommendations for both practitioners and researchers.

\subsection{Key Insights: From Inconsistent Scores to Practical Failures}

 \subsubsection{A Crisis of Consistency - The Lack of a Shared Risk Model:} 
This is not merely a technical discrepancy but points to a deeper, conceptual problem: there is no shared conceptual model of vulnerability risk among the creators and users of these systems. Each system optimizes for a different definition of risk—be it static severity (CVSS), predicted likelihood of exploit (EPSS), or stakeholder-specific impact (SSVC). This fundamental disagreement means that simply combining scores is not a viable solution, as it risks amplifying noise rather than creating clarity.
These distinct questions—``How dangerous is it?'', ``Will it be exploited soon?'', and ``What should we do now?''—naturally lead to low correlation, as a vulnerability may score high in severity but low in predicted exploitation, creating conflicting signals for prioritization. Additionally, these questions are often not sufficiently clear with respect to vulnerability management tasks, leading to misunderstandings. As a result, practitioners are left to reconcile conflicting recommendations without sufficient guidance, making consistent prioritization a moving target \textbf{(RQ1)}.

\subsubsection{The Failure of Abstraction - Triage in Theory vs. Practice}
Our analysis shows that even when used individually, scoring systems often fail in their primary practical purpose: to provide a clear, differentiated, and actionable priority list (\textbf{RQ2}). This failure leads to a significant loss of practical utility, leaving security teams with broad, undifferentiated bins that offer little actionable guidance. This problem can cause triage bottlenecks~\cite{triagebottlenecks} and alert fatigue~\cite{AlertFatigue}, effectively undermining the purpose of using a scoring system. Furthermore, the lack of consistent scoring patterns based on CWE (RQ4) suggests that high-level abstractions about vulnerability types are insufficient for guiding real-world prioritization.

\subsubsection{The Prediction Paradox - The Difficulty of Foreseeing Exploitation}
Our findings reveal a ``prediction paradox'' regarding the difficulty of foreseeing exploitation (\textbf{RQ3}). The Exploit Prediction Scoring System (EPSS), the only system designed specifically to predict future exploitation, rarely did so with high confidence before an exploit was known. This highlights the immense difficulty of real-time threat forecasting and suggests that current predictive models have limited reliability as standalone prioritization tools. It also reveals a crucial distinction: a high severity score (like from CVSS) and a high likelihood of future exploitation score (like from EPSS) are not interchangeable concepts, yet they are often treated as such in practice.

\subsection{Recommendations for Practitioners}
The evidence of systemic disagreement and practical limitations means that scoring systems should be used as carefully calibrated tools, not as absolute arbiters of risk.

\subsubsection{Treat Scores as Divergent, Advisory Inputs}
Our findings show that the perceived severity and urgency of a vulnerability can change dramatically depending on the tool used. This  complicates any attempt to create a unified and meaningful triage strategy. Therefore, practitioners should treat scores not as prescriptive commands but as advisory inputs to a broader, context-aware decision-making process. No single score provides a comprehensive view of risk.

\subsubsection{Distinguish Between Severity and Exploit Likelihood} A critical error is to treat a high severity score (like from CVSS) as being interchangeable with a high likelihood of exploitation score (like from EPSS, or KEV). Our findings caution against over-relying on EPSS for predicting future exploits, as it failed to provide a high-confidence warning for over 80\% of known exploited vulnerabilities before they were publicly disclosed.

\subsubsection{Favor Internal Contextual Augmentation Over Score Aggregation} Simply combining or averaging scores from different systems is not a solution and may amplify confusing signals rather than create clarity. A more effective approach is to use internal overlays or heuristics that interpret scores within an organization's specific context, considering factors like asset criticality and business impact.

\subsubsection{Mitigate Triage Bottlenecks from Coarse Bins} Practitioners should be aware that systems like CVSS/SSVC frequently cluster large numbers of vulnerabilities at the same severity level, creating triage bottlenecks. Organizations should develop secondary criteria or use internal risk models to further differentiate priorities within these large, overly inclusive bins.

\subsection{ Recommendations for the Research Community}
Our study reveals several key gaps in the current landscape of vulnerability scoring, pointing to important directions for future work.

\subsubsection{The Need for Explainable and Interpretable Models} The conflicting signals produced by current systems highlight a need for more explainable vulnerability scoring models. Future systems should provide transparency into the reasoning behind their scores to help users build trust and better integrate them into their decision processes.

\subsubsection{The Need for Richer Ground-Truth Severity Scores} Our study, like others, relied on proxies such as the CISA KEV catalog for ground truth on exploitation. However, these proxies are inherently incomplete. A significant research barrier to advancing this topic and developing more accurate systems is the lack of comprehensive ground-truth severity datasets. Future research should focus on creating and curating richer datasets that track not only exploitation but also capture contextual details about the actual impact on affected organizations.

\subsubsection{Rethinking Severity Scoring: Research Directions for Integrated, Context-Aware Prioritization}
Our findings show that a single, abstract severity score is insufficient to support the full range of tasks that operational vulnerability management programs require. In practice such tasks span from data intake and context enrichment through triage, remediation planning, mitigation, and governance. This points to a critical research direction: advancing beyond existing single-framework approaches by developing integrated, task-specific systems. Such systems should be designed to (i) fuse multiple scoring and context sources (e.g., CVSS, Exploitability Index, KEV, vendor advisories, and internal asset signals), (ii) produce task-tailored outputs aligned with the operational realities of each stage in the workflow, and (iii) be empirically evaluated against operational metrics such as time-to-remediation, workload reduction, and exploit capture rate. Grounding future research in this full-lifecycle, multi-signal paradigm would move vulnerability scoring from an isolated decision point toward a continuous, context-aware support system that directly improves the efficiency and impact of real-world vulnerability management programs.

\section{Related Work}
\label{sec-related-work}
A small but growing body of work compares \emph{vulnerability scoring or prioritization} approaches; however, most of these studies are \textit{qualitative}, documentation-based, or analyze a \textit{single} system in isolation rather than comparing multiple systems on the same corpus of real-world CVEs.

\subsubsection{Landscape and qualitative comparisons.}
Milousi et al.\ provide a comprehensive landscape review of scoring methodologies (CVSS, CWSS, MVSS, VIEWSS, etc.), detailing metric groups, formulas, and stated strengths/weaknesses \emph{as specified by their designers}, but do not evaluate systems head-to-head on a  dataset~\cite{milousi2024evaluating}. 
Similarly, surveys and position papers question CVSS’s suitability for prioritization~\cite{spring2021time,howland2023cvss,spring2018towards-improving} and describe SSVC’s decision-oriented goals~\cite{ssvc,ssvc-guide-cisa}, yet remain largely conceptual. 
Le et al.’s survey catalogs data-driven assessment and prioritization methods (e.g., exploit prediction, severity modeling), but focuses on technique classes rather than empirical cross-system behavior on common CVEs~\cite{le2022survey-assessment-prioritization}.

\subsubsection{Inter-rater and source inconsistency.}
Multiple works examine inconsistency \emph{within} CVSS rather than between systems: inter-rater variability when experts score the same vulnerability~\cite{holm2015expert,croft2022investigation} and skew/ambiguity in specifications~\cite{spring2021time,howland2023cvss}. 
Other studies compare \emph{data sources} (e.g., NVD vs.\ other repositories) and document metadata inconsistencies rather than operational outcomes~\cite{jiang2021evaluating,aranovich2021beyond}. 
These results motivate caution in using any single source for CVSS, but do not provide outcome-linked, cross scoring system comparisons.

\subsubsection{Outcome-linked studies focus on single signals, not cross scoring system comparisons.}
Early empirical work relates severity to exploitation using case-control designs~\cite{allodi2014comparing} and Bayesian analysis of CVSS trustworthiness~\cite{7797152}. 
EPSS is developed as a prediction signal using historical exploitation data~\cite{jacobs2021exploit,jacobs2020improving}, while later work models exploit \emph{development} likelihood~\cite{suciu2022expected}. 
These studies are outcome-linked but typically evaluate \textit{one} scoring/prediction signal at a time (e.g., EPSS), as opposed to a side-by-side comparison of heterogeneous systems (severity, threat-likelihood, and decision-support) applied to the same CVEs.

\subsubsection{Context-aware prioritization and CVSS-centric enhancements.}
A large stream of work augments CVSS (re-weighting, temporal, environmental variants, text mining/ML scoring)~\cite{huang2013novel,wivss,liu2011vrss,liu2012improving,text-mining,fruhwirth2009improving,jung2022cavp,walkowski2021automatic,walkowski2021vulnerability,figueroa2020survey}. 
Context-aware frameworks (e.g., Vulcon, Vulman, and related models) integrate asset criticality, mission impact, inventories, or reinforcement learning for selection/triage~\cite{vulcon,vulman,ahmadi2022automated,hore2022towards,shah2022math,bulut2022vulnerability,reyes2022environment}, but still \emph{consume} CVSS or similar signals as inputs and report case-specific evaluations rather than cross-system, outcome-linked comparisons. 
Domain/proprietary systems (e.g., Microsoft Exploitability Index, Tenable VPR, Qualys VMDR, Rapid7 Nexpose) extend inputs with threat/asset context~\cite{microsoft-exploitability,vpr1,vpr2,vmdr,nexpose,recorded-future,snyk}, yet are either ecosystem-specific or opaque, limiting reproducible comparative studies.

\subsubsection{Gap and positioning.}
Across these threads, we find (i) few \emph{comparative} studies of multiple scoring systems on the \emph{same, operationally relevant} CVE set; (ii) limited \emph{outcome-linked} evaluation that aligns system outputs with real-world exploitation (e.g., KEV) \emph{across} systems; and (iii) little analysis of how differing goals (severity vs.\ threat likelihood vs.\ action recommendation) and input types (static vs.\ dynamic; context-free vs.\ context-aware) drive disagreement in practice. 
To our knowledge, prior work has not provided a large-scale, \emph{industry-grounded} empirical comparison of CVSS, EPSS, SSVC, and Exploitability Index applied to a shared Patch Tuesday corpus, with quantitative measures of inter-system agreement, triage burden, and exploitation alignment~\cite{ssvc,epss,epss-paper,microsoft-exploitability}. 
Our study addresses this gap by supplying reproducible, cross-system, outcome-linked evidence that complements (and extends beyond) prior qualitative and single-signal analyses.

\section{Conclusions}
\label{sec:conclusion}

This paper presents the first large-scale, empirical evaluation of four prominent vulnerability scoring systems—CVSS, EPSS, SSVC, and the Exploitability Index—using a real-world dataset of 600 vulnerabilities from Microsoft's Patch Tuesday disclosures. Our study was designed to fill a critical gap left by prior work, which has been largely qualitative, by providing quantitative evidence of how these systems perform in an operational context. The findings demonstrate considerable and systemic disagreement among the systems, which exhibit little to no correlation or categorical agreement when scoring the same vulnerabilities. We found that all four systems produce overly broad priority groups that complicate triage efforts and that predictive systems like EPSS often fail to flag known exploited vulnerabilities ahead of time, with fewer than 20\% of CISA KEV CVEs receiving a high-confidence score before exploitation was public.

The central implication of this research is that these widely used scoring systems are not interchangeable and their conflicting guidance reveals a deeper, systemic issue: a lack of a shared conceptual model of risk across the vulnerability management ecosystem. The observed divergence is a direct result of each system's unique design goals—measuring inherent severity versus predicting threat likelihood versus recommending a specific action. Given these findings, we caution practitioners against relying on any one system as the sole basis for prioritization; scores should be treated as advisory inputs to a broader, context-aware process. Ultimately, our study highlights an urgent need for the research community to develop more transparent, interpretable, and task-specific frameworks that are empirically grounded and better aligned with the practical realities of cybersecurity operations.





\bibliographystyle{ACM-Reference-Format} 
\bibliography{references}


\begin{thebibliography}{68}


\ifx \showCODEN    \undefined \def \showCODEN     #1{\unskip}     \fi
\ifx \showISBNx    \undefined \def \showISBNx     #1{\unskip}     \fi
\ifx \showISBNxiii \undefined \def \showISBNxiii  #1{\unskip}     \fi
\ifx \showISSN     \undefined \def \showISSN      #1{\unskip}     \fi
\ifx \showLCCN     \undefined \def \showLCCN      #1{\unskip}     \fi
\ifx \shownote     \undefined \def \shownote      #1{#1}          \fi
\ifx \showarticletitle \undefined \def \showarticletitle #1{#1}   \fi
\ifx \showURL      \undefined \def \showURL       {\relax}        \fi
\providecommand\bibfield[2]{#2}
\providecommand\bibinfo[2]{#2}
\providecommand\natexlab[1]{#1}
\providecommand\showeprint[2][]{arXiv:#2}

\bibitem[Ahmadi~Mehri et~al\mbox{.}(2022)]%
        {ahmadi2022automated}
\bibfield{author}{\bibinfo{person}{Vida Ahmadi~Mehri}, \bibinfo{person}{Patrik Arlos}, {and} \bibinfo{person}{Emiliano Casalicchio}.} \bibinfo{year}{2022}\natexlab{}.
\newblock \showarticletitle{Automated Context-Aware Vulnerability Risk Management for Patch Prioritization}.
\newblock \bibinfo{journal}{\emph{Electronics}} \bibinfo{volume}{11}, \bibinfo{number}{21} (\bibinfo{year}{2022}), \bibinfo{pages}{3580}.
\newblock


\bibitem[Allodi and Massacci(2014)]%
        {allodi2014comparing}
\bibfield{author}{\bibinfo{person}{Luca Allodi} {and} \bibinfo{person}{Fabio Massacci}.} \bibinfo{year}{2014}\natexlab{}.
\newblock \showarticletitle{Comparing vulnerability severity and exploits using case-control studies}.
\newblock \bibinfo{journal}{\emph{ACM Transactions on Information and System Security (TISSEC)}} \bibinfo{volume}{17}, \bibinfo{number}{1} (\bibinfo{year}{2014}), \bibinfo{pages}{1--20}.
\newblock


\bibitem[Aranovich et~al\mbox{.}(2021)]%
        {aranovich2021beyond}
\bibfield{author}{\bibinfo{person}{Ra{\'u}l Aranovich}, \bibinfo{person}{Muting Wu}, \bibinfo{person}{Dian Yu}, \bibinfo{person}{Katya Katsy}, \bibinfo{person}{Benyamin Ahmadnia}, \bibinfo{person}{Matthew Bishop}, \bibinfo{person}{Vladimir Filkov}, {and} \bibinfo{person}{Kenji Sagae}.} \bibinfo{year}{2021}\natexlab{}.
\newblock \showarticletitle{Beyond NVD: Cybersecurity meets the Semantic Web.}. In \bibinfo{booktitle}{\emph{New Security Paradigms Workshop}}. \bibinfo{pages}{59--69}.
\newblock


\bibitem[Booth et~al\mbox{.}(2013)]%
        {nvd}
\bibfield{author}{\bibinfo{person}{Harold Booth}, \bibinfo{person}{Doug Rike}, {and} \bibinfo{person}{Gregory~A Witte}.} \bibinfo{year}{2013}\natexlab{}.
\newblock \showarticletitle{The national vulnerability database (nvd): Overview}.
\newblock  (\bibinfo{year}{2013}).
\newblock


\bibitem[Bulut et~al\mbox{.}(2022)]%
        {bulut2022vulnerability}
\bibfield{author}{\bibinfo{person}{Muhammed~Fatih Bulut}, \bibinfo{person}{Abdulhamid Adebayo}, \bibinfo{person}{Daby Sow}, {and} \bibinfo{person}{Steve Ocepek}.} \bibinfo{year}{2022}\natexlab{}.
\newblock \showarticletitle{Vulnerability prioritization: An offensive security approach}.
\newblock \bibinfo{journal}{\emph{arXiv preprint arXiv:2206.11182}} (\bibinfo{year}{2022}).
\newblock


\bibitem[Croft et~al\mbox{.}(2022)]%
        {croft2022investigation}
\bibfield{author}{\bibinfo{person}{Roland Croft}, \bibinfo{person}{M~Ali Babar}, {and} \bibinfo{person}{Li Li}.} \bibinfo{year}{2022}\natexlab{}.
\newblock \showarticletitle{An investigation into inconsistency of software vulnerability severity across data sources}. In \bibinfo{booktitle}{\emph{2022 IEEE International Conference on Software Analysis, Evolution and Reengineering (SANER)}}. IEEE, \bibinfo{pages}{338--348}.
\newblock


\bibitem[Cybersecurity and Agency(2022)]%
        {ssvc-guide-cisa}
\bibfield{author}{\bibinfo{person}{Cybersecurity} {and} \bibinfo{person}{Infrastructure~Security Agency}.} \bibinfo{year}{2022}\natexlab{}.
\newblock \bibinfo{booktitle}{\emph{CISA Stakeholder-Specific Vulnerability Categorization Guide}}.
\newblock


\bibitem[{Cybersecurity and Infrastructure Security Agency}({[n.\,d.]})]%
        {cisa-website-ssvc}
\bibfield{author}{\bibinfo{person}{{Cybersecurity and Infrastructure Security Agency}}.} \bibinfo{year}{[n.\,d.]}\natexlab{}.
\newblock \bibinfo{title}{{Stakeholder-Specific Vulnerability Categorization (SSVC)}}.
\newblock \bibinfo{howpublished}{\url{https://www.cisa.gov/stakeholder-specific-vulnerability-categorization-ssvc}}.
\newblock


\bibitem[Dugal and Rich(2023)]%
        {cvss4}
\bibfield{author}{\bibinfo{person}{Dave Dugal} {and} \bibinfo{person}{Dale Rich}.} \bibinfo{year}{2023}\natexlab{}.
\newblock \bibinfo{title}{{Announcing CVSS v4.0}}.
\newblock \bibinfo{howpublished}{35th Annual FIRST Conference}.
\newblock
\newblock
\shownote{Available: \url{https://www.first.org/cvss/v4-0/cvss-v40-presentation.pdf}}.


\bibitem[Farris et~al\mbox{.}(2018)]%
        {vulcon}
\bibfield{author}{\bibinfo{person}{Katheryn~A Farris}, \bibinfo{person}{Ankit Shah}, \bibinfo{person}{George Cybenko}, \bibinfo{person}{Rajesh Ganesan}, {and} \bibinfo{person}{Sushil Jajodia}.} \bibinfo{year}{2018}\natexlab{}.
\newblock \showarticletitle{Vulcon: A system for vulnerability prioritization, mitigation, and management}.
\newblock \bibinfo{journal}{\emph{ACM Transactions on Privacy and Security (TOPS)}} \bibinfo{volume}{21}, \bibinfo{number}{4} (\bibinfo{year}{2018}), \bibinfo{pages}{1--28}.
\newblock


\bibitem[Figueroa-Lorenzo et~al\mbox{.}(2020)]%
        {figueroa2020survey}
\bibfield{author}{\bibinfo{person}{Santiago Figueroa-Lorenzo}, \bibinfo{person}{Javier A{\~n}orga}, {and} \bibinfo{person}{Saioa Arrizabalaga}.} \bibinfo{year}{2020}\natexlab{}.
\newblock \showarticletitle{A survey of IIoT protocols: A measure of vulnerability risk analysis based on CVSS}.
\newblock \bibinfo{journal}{\emph{ACM Computing Surveys (CSUR)}} \bibinfo{volume}{53}, \bibinfo{number}{2} (\bibinfo{year}{2020}), \bibinfo{pages}{1--53}.
\newblock


\bibitem[{FIRST.Org, Inc.}({[n.\,d.]})]%
        {cvss4website}
\bibfield{author}{\bibinfo{person}{{FIRST.Org, Inc.}}} \bibinfo{year}{[n.\,d.]}\natexlab{}.
\newblock \bibinfo{title}{Common Vulnerability Scoring System Version 4.0}.
\newblock \bibinfo{howpublished}{\url{https://www.first.org/cvss/v4-0/}}.
\newblock


\bibitem[{FIRST.Org, Inc.}(2019)]%
        {cvss3.1}
\bibfield{author}{\bibinfo{person}{{FIRST.Org, Inc.}}} \bibinfo{year}{2019}\natexlab{}.
\newblock \bibinfo{booktitle}{\emph{Common Vulnerability Scoring System version 3.1 Specification Document Revision 1}}.
\newblock \bibinfo{type}{{T}echnical {R}eport}. \bibinfo{institution}{{FIRST.Org, Inc.}}
\newblock
\urldef\tempurl%
\url{https://www.first.org/cvss/v3-1/cvss-v31-specification\_r1.pdf}
\showURL{%
\tempurl}


\bibitem[{First.org, Inc.}(2023)]%
        {epss2}
\bibfield{author}{\bibinfo{person}{{First.org, Inc.}}} \bibinfo{year}{2023}\natexlab{}.
\newblock \bibinfo{title}{{EPSS Frequently Asked Questions}}.
\newblock \bibinfo{howpublished}{\url{https://www.first.org/epss/faq}}.
\newblock


\bibitem[{First.org, Inc.}(2025)]%
        {epss}
\bibfield{author}{\bibinfo{person}{{First.org, Inc.}}} \bibinfo{year}{2025}\natexlab{}.
\newblock \bibinfo{title}{{Exploit Prediction Scoring System}}.
\newblock \bibinfo{howpublished}{\url{https://www.first.org/epss/}}.
\newblock


\bibitem[for Disease~Control and Prevention(2019)]%
        {cdc}
\bibfield{author}{\bibinfo{person}{Centers for Disease~Control} {and} \bibinfo{person}{Prevention}.} \bibinfo{year}{2019}\natexlab{}.
\newblock \bibinfo{title}{How is well-being defined?}
\newblock \bibinfo{howpublished}{\url{https://www.cdc.gov/hrqol/wellbeing.htm\#three}}.
\newblock


\bibitem[Foreman(2019)]%
        {foreman2019vulnerability}
\bibfield{author}{\bibinfo{person}{Park Foreman}.} \bibinfo{year}{2019}\natexlab{}.
\newblock \bibinfo{booktitle}{\emph{Vulnerability management}}.
\newblock \bibinfo{publisher}{CRC Press}.
\newblock


\bibitem[Fruhwirth and Mannisto(2009)]%
        {fruhwirth2009improving}
\bibfield{author}{\bibinfo{person}{Christian Fruhwirth} {and} \bibinfo{person}{Tomi Mannisto}.} \bibinfo{year}{2009}\natexlab{}.
\newblock \showarticletitle{Improving CVSS-based vulnerability prioritization and response with context information}. In \bibinfo{booktitle}{\emph{2009 3rd International symposium on empirical software engineering and measurement}}. IEEE, \bibinfo{pages}{535--544}.
\newblock


\bibitem[Hans and Brandtweiner(2022)]%
        {hans2022best}
\bibfield{author}{\bibinfo{person}{Jaqueline Hans} {and} \bibinfo{person}{Roman Brandtweiner}.} \bibinfo{year}{2022}\natexlab{}.
\newblock \showarticletitle{BEST PRACTICES FOR VULNERABILITY MANAGEMENT IN LARGE ENTERPRISES: A CRITICAL VIEW ON THE COMMON VULNERABILITY SCORING SYSTEM}.
\newblock \bibinfo{journal}{\emph{Risk Analysis, Hazard Mitigation and Safety and Security Engineering XIII}}  \bibinfo{volume}{214} (\bibinfo{year}{2022}), \bibinfo{pages}{123}.
\newblock


\bibitem[Holm and Afridi(2015)]%
        {holm2015expert}
\bibfield{author}{\bibinfo{person}{Hannes Holm} {and} \bibinfo{person}{Khalid~Khan Afridi}.} \bibinfo{year}{2015}\natexlab{}.
\newblock \showarticletitle{An expert-based investigation of the common vulnerability scoring system}.
\newblock \bibinfo{journal}{\emph{Computers \& Security}}  \bibinfo{volume}{53} (\bibinfo{year}{2015}), \bibinfo{pages}{18--30}.
\newblock


\bibitem[Hore et~al\mbox{.}(2022)]%
        {hore2022towards}
\bibfield{author}{\bibinfo{person}{Soumyadeep Hore}, \bibinfo{person}{Fariha Moomtaheen}, \bibinfo{person}{Ankit Shah}, {and} \bibinfo{person}{Xinming Ou}.} \bibinfo{year}{2022}\natexlab{}.
\newblock \showarticletitle{Towards optimal triage and mitigation of context-sensitive cyber vulnerabilities}.
\newblock \bibinfo{journal}{\emph{IEEE Transactions on Dependable and Secure Computing}} \bibinfo{volume}{20}, \bibinfo{number}{2} (\bibinfo{year}{2022}), \bibinfo{pages}{1270--1285}.
\newblock


\bibitem[Hore et~al\mbox{.}(2023)]%
        {vulman}
\bibfield{author}{\bibinfo{person}{Soumyadeep Hore}, \bibinfo{person}{Ankit Shah}, {and} \bibinfo{person}{Nathaniel~D Bastian}.} \bibinfo{year}{2023}\natexlab{}.
\newblock \showarticletitle{Deep VULMAN: A deep reinforcement learning-enabled cyber vulnerability management framework}.
\newblock \bibinfo{journal}{\emph{Expert Systems with Applications}}  \bibinfo{volume}{221} (\bibinfo{year}{2023}), \bibinfo{pages}{119734}.
\newblock


\bibitem[Howland(2023)]%
        {howland2023cvss}
\bibfield{author}{\bibinfo{person}{Henry Howland}.} \bibinfo{year}{2023}\natexlab{}.
\newblock \showarticletitle{Cvss: Ubiquitous and broken}.
\newblock \bibinfo{journal}{\emph{Digital Threats: Research and Practice}} \bibinfo{volume}{4}, \bibinfo{number}{1} (\bibinfo{year}{2023}), \bibinfo{pages}{1--12}.
\newblock


\bibitem[Huang et~al\mbox{.}(2013)]%
        {huang2013novel}
\bibfield{author}{\bibinfo{person}{Chien-Cheng Huang}, \bibinfo{person}{Feng-Yu Lin}, \bibinfo{person}{Frank Yeong-Sung Lin}, {and} \bibinfo{person}{Yeali~S Sun}.} \bibinfo{year}{2013}\natexlab{}.
\newblock \showarticletitle{A novel approach to evaluate software vulnerability prioritization}.
\newblock \bibinfo{journal}{\emph{Journal of Systems and Software}} \bibinfo{volume}{86}, \bibinfo{number}{11} (\bibinfo{year}{2013}), \bibinfo{pages}{2822--2840}.
\newblock


\bibitem[Institute(2020)]%
        {Ponemon2020VulnMgmt}
\bibfield{author}{\bibinfo{person}{Ponemon Institute}.} \bibinfo{year}{2020}\natexlab{}.
\newblock \bibinfo{title}{Ponemon Study on the Challenging State of Vulnerability Management}.
\newblock \bibinfo{howpublished}{\url{https://www.balbix.com/press-releases/ponemon-report-on-vulnerability-management-challenges/}}.
\newblock
\newblock
\shownote{Accessed: 2025-07-14}.


\bibitem[Jacobs et~al\mbox{.}(2020)]%
        {jacobs2020improving}
\bibfield{author}{\bibinfo{person}{Jay Jacobs}, \bibinfo{person}{Sasha Romanosky}, \bibinfo{person}{Idris Adjerid}, {and} \bibinfo{person}{Wade Baker}.} \bibinfo{year}{2020}\natexlab{}.
\newblock \showarticletitle{Improving vulnerability remediation through better exploit prediction}.
\newblock \bibinfo{journal}{\emph{Journal of Cybersecurity}} \bibinfo{volume}{6}, \bibinfo{number}{1} (\bibinfo{year}{2020}), \bibinfo{pages}{tyaa015}.
\newblock


\bibitem[Jacobs et~al\mbox{.}(2021)]%
        {jacobs2021exploit}
\bibfield{author}{\bibinfo{person}{Jay Jacobs}, \bibinfo{person}{Sasha Romanosky}, \bibinfo{person}{Benjamin Edwards}, \bibinfo{person}{Idris Adjerid}, {and} \bibinfo{person}{Michael Roytman}.} \bibinfo{year}{2021}\natexlab{}.
\newblock \showarticletitle{Exploit prediction scoring system (epss)}.
\newblock \bibinfo{journal}{\emph{Digital Threats: Research and Practice}} \bibinfo{volume}{2}, \bibinfo{number}{3} (\bibinfo{year}{2021}), \bibinfo{pages}{1--17}.
\newblock


\bibitem[Jacobs et~al\mbox{.}(2023)]%
        {epss-paper}
\bibfield{author}{\bibinfo{person}{Jay Jacobs}, \bibinfo{person}{Sasha Romanosky}, \bibinfo{person}{Octavian Suciu}, \bibinfo{person}{Ben Edwards}, {and} \bibinfo{person}{Armin Sarabi}.} \bibinfo{year}{2023}\natexlab{}.
\newblock \showarticletitle{Enhancing Vulnerability prioritization: Data-driven exploit predictions with community-driven insights}. In \bibinfo{booktitle}{\emph{2023 IEEE European Symposium on Security and Privacy Workshops (EuroS\&PW)}}. IEEE, \bibinfo{pages}{194--206}.
\newblock


\bibitem[Jiang et~al\mbox{.}(2021)]%
        {jiang2021evaluating}
\bibfield{author}{\bibinfo{person}{Yuning Jiang}, \bibinfo{person}{Manfred Jeusfeld}, {and} \bibinfo{person}{Jianguo Ding}.} \bibinfo{year}{2021}\natexlab{}.
\newblock \showarticletitle{Evaluating the data inconsistency of open-source vulnerability repositories}. In \bibinfo{booktitle}{\emph{Proceedings of the 16th International Conference on Availability, Reliability and Security}}. \bibinfo{pages}{1--10}.
\newblock


\bibitem[Johnson et~al\mbox{.}(2018)]%
        {7797152}
\bibfield{author}{\bibinfo{person}{Pontus Johnson}, \bibinfo{person}{Robert Lagerström}, \bibinfo{person}{Mathias Ekstedt}, {and} \bibinfo{person}{Ulrik Franke}.} \bibinfo{year}{2018}\natexlab{}.
\newblock \showarticletitle{Can the Common Vulnerability Scoring System be Trusted? A Bayesian Analysis}.
\newblock \bibinfo{journal}{\emph{IEEE Transactions on Dependable and Secure Computing}} \bibinfo{volume}{15}, \bibinfo{number}{6} (\bibinfo{year}{2018}), \bibinfo{pages}{1002--1015}.
\newblock
\href{https://doi.org/10.1109/TDSC.2016.2644614}{doi:\nolinkurl{10.1109/TDSC.2016.2644614}}


\bibitem[Jung et~al\mbox{.}(2022)]%
        {jung2022cavp}
\bibfield{author}{\bibinfo{person}{Bill Jung}, \bibinfo{person}{Yan Li}, {and} \bibinfo{person}{Tamir Bechor}.} \bibinfo{year}{2022}\natexlab{}.
\newblock \showarticletitle{CAVP: A context-aware vulnerability prioritization model}.
\newblock \bibinfo{journal}{\emph{Computers \& Security}}  \bibinfo{volume}{116} (\bibinfo{year}{2022}), \bibinfo{pages}{102639}.
\newblock


\bibitem[Kossakowski et~al\mbox{.}(2019)]%
        {kossakowski2019computer}
\bibfield{author}{\bibinfo{person}{Klaus-Peter Kossakowski}, \bibinfo{person}{Vilius Benetis}, \bibinfo{person}{Olivier Caleff}, \bibinfo{person}{Cristine Hoepers}, \bibinfo{person}{Angela Horneman}, \bibinfo{person}{Allen Householder}, \bibinfo{person}{Art Manion}, \bibinfo{person}{Amanda Mullens}, \bibinfo{person}{Samuel Perl}, \bibinfo{person}{Daniel Roethlisberger}, \bibinfo{person}{Sigitas Rokas}, \bibinfo{person}{Mary Rossell}, \bibinfo{person}{Robin~M. Ruefle}, \bibinfo{person}{Désirée Sacher}, \bibinfo{person}{Krassimir~T. Tzvetanov}, {and} \bibinfo{person}{Mark Zajicek}.} \bibinfo{year}{2019}\natexlab{}.
\newblock \bibinfo{booktitle}{\emph{Computer Security Incident Response Team (CSIRT) Services Framework version 2.1.0}}.
\newblock \bibinfo{type}{{T}echnical {R}eport}. \bibinfo{institution}{FIRST.Org, Inc.}
\newblock


\bibitem[Krippendorff(2022)]%
        {krippendorff2022content}
\bibfield{author}{\bibinfo{person}{Klaus Krippendorff}.} \bibinfo{year}{2022}\natexlab{}.
\newblock \bibinfo{booktitle}{\emph{Content Analysis: An Introduction to Its Methodology} (\bibinfo{edition}{4th} ed.)}.
\newblock \bibinfo{publisher}{SAGE Publications}, \bibinfo{address}{Thousand Oaks, CA}.
\newblock
\showISBNx{9781544391026}


\bibitem[Le et~al\mbox{.}(2022)]%
        {le2022survey-assessment-prioritization}
\bibfield{author}{\bibinfo{person}{Triet~HM Le}, \bibinfo{person}{Huaming Chen}, {and} \bibinfo{person}{M~Ali Babar}.} \bibinfo{year}{2022}\natexlab{}.
\newblock \showarticletitle{A survey on data-driven software vulnerability assessment and prioritization}.
\newblock \bibinfo{journal}{\emph{Comput. Surveys}} \bibinfo{volume}{55}, \bibinfo{number}{5} (\bibinfo{year}{2022}), \bibinfo{pages}{1--39}.
\newblock


\bibitem[Liu et~al\mbox{.}(2022)]%
        {AlertFatigue}
\bibfield{author}{\bibinfo{person}{Jia Liu}, \bibinfo{person}{Runzi Zhang}, \bibinfo{person}{Wenmao Liu}, \bibinfo{person}{Yinghua Zhang}, \bibinfo{person}{Dujuan Gu}, \bibinfo{person}{Mingkai Tong}, \bibinfo{person}{Xingkai Wang}, \bibinfo{person}{Jianxin Xue}, {and} \bibinfo{person}{Huanran Wang}.} \bibinfo{year}{2022}\natexlab{}.
\newblock \showarticletitle{Context2Vector: Accelerating security event triage via context representation learning}.
\newblock \bibinfo{journal}{\emph{Information and Software Technology}}  \bibinfo{volume}{146} (\bibinfo{year}{2022}), \bibinfo{pages}{106856}.
\newblock
\showISSN{0950-5849}
\href{https://doi.org/10.1016/j.infsof.2022.106856}{doi:\nolinkurl{10.1016/j.infsof.2022.106856}}


\bibitem[Liu and Zhang(2011)]%
        {liu2011vrss}
\bibfield{author}{\bibinfo{person}{Qixu Liu} {and} \bibinfo{person}{Yuqing Zhang}.} \bibinfo{year}{2011}\natexlab{}.
\newblock \showarticletitle{VRSS: A new system for rating and scoring vulnerabilities}.
\newblock \bibinfo{journal}{\emph{Computer Communications}} \bibinfo{volume}{34}, \bibinfo{number}{3} (\bibinfo{year}{2011}), \bibinfo{pages}{264--273}.
\newblock


\bibitem[Liu et~al\mbox{.}(2012)]%
        {liu2012improving}
\bibfield{author}{\bibinfo{person}{Qixu Liu}, \bibinfo{person}{Yuqing Zhang}, \bibinfo{person}{Ying Kong}, {and} \bibinfo{person}{Qianru Wu}.} \bibinfo{year}{2012}\natexlab{}.
\newblock \showarticletitle{Improving VRSS-based vulnerability prioritization using analytic hierarchy process}.
\newblock \bibinfo{journal}{\emph{Journal of Systems and Software}} \bibinfo{volume}{85}, \bibinfo{number}{8} (\bibinfo{year}{2012}), \bibinfo{pages}{1699--1708}.
\newblock


\bibitem[{Lockheed Martin}({[n.\,d.]})]%
        {cyber-kill-chain}
\bibfield{author}{\bibinfo{person}{{Lockheed Martin}}.} \bibinfo{year}{[n.\,d.]}\natexlab{}.
\newblock \bibinfo{title}{{The Cyber Kill Chain}}.
\newblock \bibinfo{howpublished}{\url{https://www.lockheedmartin.com/en-us/capabilities/cyber/cyber-kill-chain.html}}.
\newblock


\bibitem[{Microsoft, Inc.}({[n.\,d.]})]%
        {microsoft-exploitability}
\bibfield{author}{\bibinfo{person}{{Microsoft, Inc.}}} \bibinfo{year}{[n.\,d.]}\natexlab{}.
\newblock \bibinfo{title}{{Microsoft Exploitability Index}}.
\newblock \bibinfo{howpublished}{\url{https://www.microsoft.com/en-us/msrc/exploitability-index/}}.
\newblock


\bibitem[Milousi et~al\mbox{.}(2024)]%
        {milousi2024evaluating}
\bibfield{author}{\bibinfo{person}{Konstantina Milousi}, \bibinfo{person}{Prodromos Kiriakidis}, \bibinfo{person}{Notis Mengidis}, \bibinfo{person}{Georgios Rizos}, \bibinfo{person}{Mariana~S Mazi}, \bibinfo{person}{Antonis Voulgaridis}, \bibinfo{person}{Konstantinos Votis}, {and} \bibinfo{person}{Dimitrios Tzovaras}.} \bibinfo{year}{2024}\natexlab{}.
\newblock \showarticletitle{Evaluating Cybersecurity Risk: A Comprehensive Comparison of Vulnerability Scoring Methodologies}. In \bibinfo{booktitle}{\emph{Proceedings of the 19th International Conference on Availability, Reliability and Security}}. \bibinfo{pages}{1--11}.
\newblock


\bibitem[Okutan and Mirakhorli(2022a)]%
        {ExploitabilityIndex-github}
\bibfield{author}{\bibinfo{person}{Ahmet Okutan} {and} \bibinfo{person}{Mehdi Mirakhorli}.} \bibinfo{year}{2022}\natexlab{a}.
\newblock \bibinfo{title}{{Exploitability Analysis Datasets and Models}}.
\newblock \bibinfo{howpublished}{\url{https://github.com/SoftwareDesignLab/exploitability_analysis}}.
\newblock


\bibitem[Okutan and Mirakhorli(2022b)]%
        {exp_index_2022}
\bibfield{author}{\bibinfo{person}{Ahmet Okutan} {and} \bibinfo{person}{Mehdi Mirakhorli}.} \bibinfo{year}{2022}\natexlab{b}.
\newblock \showarticletitle{Predicting the severity and exploitability of vulnerability reports using convolutional neural nets}. In \bibinfo{booktitle}{\emph{Proceedings of the 3rd International Workshop on Engineering and Cybersecurity of Critical Systems}} (Pittsburgh, Pennsylvania) \emph{(\bibinfo{series}{EnCyCriS '22})}. \bibinfo{publisher}{Association for Computing Machinery}, \bibinfo{address}{New York, NY, USA}, \bibinfo{pages}{1–8}.
\newblock
\showISBNx{9781450392907}
\href{https://doi.org/10.1145/3524489.3527298}{doi:\nolinkurl{10.1145/3524489.3527298}}


\bibitem[Puth et~al\mbox{.}(2015)]%
        {puth2015effective}
\bibfield{author}{\bibinfo{person}{Marie-Therese Puth}, \bibinfo{person}{Markus Neuh{\"a}user}, {and} \bibinfo{person}{Graeme~D Ruxton}.} \bibinfo{year}{2015}\natexlab{}.
\newblock \showarticletitle{Effective use of Spearman's and Kendall's correlation coefficients for association between two measured traits}.
\newblock \bibinfo{journal}{\emph{Animal Behaviour}}  \bibinfo{volume}{102} (\bibinfo{year}{2015}), \bibinfo{pages}{77--84}.
\newblock


\bibitem[{QED Secure Solutions}({[n.\,d.]})]%
        {rss}
\bibfield{author}{\bibinfo{person}{{QED Secure Solutions}}.} \bibinfo{year}{[n.\,d.]}\natexlab{}.
\newblock \bibinfo{title}{{Risk Scoring System}}.
\newblock \bibinfo{howpublished}{\url{https://www.riskscoringsystem.com/}}.
\newblock


\bibitem[{Qualys, Inc.}({[n.\,d.]})]%
        {vmdr}
\bibfield{author}{\bibinfo{person}{{Qualys, Inc.}}} \bibinfo{year}{[n.\,d.]}\natexlab{}.
\newblock \bibinfo{title}{{Qualys Vulnerability Management, Detection, and Response Tool}}.
\newblock \bibinfo{howpublished}{\url{https://www.qualys.com/apps/vulnerability-management-detection-response/}}.
\newblock


\bibitem[{Rapid7, Inc.}({[n.\,d.]})]%
        {nexpose}
\bibfield{author}{\bibinfo{person}{{Rapid7, Inc.}}} \bibinfo{year}{[n.\,d.]}\natexlab{}.
\newblock \bibinfo{title}{{Nexpose Vulnerability Scanner}}.
\newblock \bibinfo{howpublished}{\url{https://www.rapid7.com/products/nexpose/}}.
\newblock


\bibitem[{Recorded Future, Inc.}({[n.\,d.]})]%
        {recorded-future}
\bibfield{author}{\bibinfo{person}{{Recorded Future, Inc.}}} \bibinfo{year}{[n.\,d.]}\natexlab{}.
\newblock \bibinfo{title}{{Recorded Future Threat Intelligence}}.
\newblock \bibinfo{howpublished}{\url{https://www.recordedfuture.com/}}.
\newblock


\bibitem[Reyes et~al\mbox{.}(2022)]%
        {reyes2022environment}
\bibfield{author}{\bibinfo{person}{Jorge Reyes}, \bibinfo{person}{Walter Fuertes}, \bibinfo{person}{Paco Ar{\'e}valo}, {and} \bibinfo{person}{Mayra Macas}.} \bibinfo{year}{2022}\natexlab{}.
\newblock \showarticletitle{An Environment-Specific Prioritization Model for Information-Security Vulnerabilities Based on Risk Factor Analysis}.
\newblock \bibinfo{journal}{\emph{Electronics}} \bibinfo{volume}{11}, \bibinfo{number}{9} (\bibinfo{year}{2022}), \bibinfo{pages}{1334}.
\newblock


\bibitem[Runeson and H{\"o}st(2009)]%
        {Runeson2009}
\bibfield{author}{\bibinfo{person}{Per Runeson} {and} \bibinfo{person}{Martin H{\"o}st}.} \bibinfo{year}{2009}\natexlab{}.
\newblock \showarticletitle{Guidelines for conducting and reporting case study research in software engineering}.
\newblock \bibinfo{journal}{\emph{Empirical Software Engineering}}  \bibinfo{volume}{14} (\bibinfo{year}{2009}), \bibinfo{pages}{131--164}.
\newblock
\urldef\tempurl%
\url{https://api.semanticscholar.org/CorpusID:207144526}
\showURL{%
\tempurl}


\bibitem[Shah et~al\mbox{.}(2022)]%
        {shah2022math}
\bibfield{author}{\bibinfo{person}{Ankit Shah}, \bibinfo{person}{Katheryn~A Farris}, \bibinfo{person}{Rajesh Ganesan}, {and} \bibinfo{person}{Sushil Jajodia}.} \bibinfo{year}{2022}\natexlab{}.
\newblock \showarticletitle{Vulnerability selection for remediation: An empirical analysis}.
\newblock \bibinfo{journal}{\emph{The Journal of Defense Modeling and Simulation}} \bibinfo{volume}{19}, \bibinfo{number}{1} (\bibinfo{year}{2022}), \bibinfo{pages}{13--22}.
\newblock


\bibitem[{Snyk, Ltd.}({[n.\,d.]})]%
        {snyk}
\bibfield{author}{\bibinfo{person}{{Snyk, Ltd.}}} \bibinfo{year}{[n.\,d.]}\natexlab{}.
\newblock \bibinfo{title}{{Snyk Priority Score}}.
\newblock \bibinfo{howpublished}{\url{https://snyk.io/}}.
\newblock


\bibitem[Spanos et~al\mbox{.}(2017)]%
        {text-mining}
\bibfield{author}{\bibinfo{person}{Georgios Spanos}, \bibinfo{person}{Lefteris Angelis}, {and} \bibinfo{person}{Dimitrios Toloudis}.} \bibinfo{year}{2017}\natexlab{}.
\newblock \showarticletitle{Assessment of vulnerability severity using text mining}. In \bibinfo{booktitle}{\emph{Proceedings of the 21st Pan-Hellenic conference on informatics}}. \bibinfo{pages}{1--6}.
\newblock


\bibitem[Spanos et~al\mbox{.}(2013)]%
        {wivss}
\bibfield{author}{\bibinfo{person}{Georgios Spanos}, \bibinfo{person}{Angeliki Sioziou}, {and} \bibinfo{person}{Lefteris Angelis}.} \bibinfo{year}{2013}\natexlab{}.
\newblock \showarticletitle{WIVSS: a new methodology for scoring information systems vulnerabilities}. In \bibinfo{booktitle}{\emph{Proceedings of the 17th panhellenic conference on informatics}}. \bibinfo{pages}{83--90}.
\newblock


\bibitem[Spring et~al\mbox{.}(2021a)]%
        {spring2021time}
\bibfield{author}{\bibinfo{person}{Jonathan Spring}, \bibinfo{person}{Eric Hatleback}, \bibinfo{person}{Allen Householder}, \bibinfo{person}{Art Manion}, {and} \bibinfo{person}{Deana Shick}.} \bibinfo{year}{2021}\natexlab{a}.
\newblock \showarticletitle{Time to Change the CVSS?}
\newblock \bibinfo{journal}{\emph{IEEE Security \& Privacy}} \bibinfo{volume}{19}, \bibinfo{number}{2} (\bibinfo{year}{2021}), \bibinfo{pages}{74--78}.
\newblock


\bibitem[Spring et~al\mbox{.}(2018)]%
        {spring2018towards-improving}
\bibfield{author}{\bibinfo{person}{Jonathan Spring}, \bibinfo{person}{Eric Hatleback}, \bibinfo{person}{A Manion}, {and} \bibinfo{person}{D Shic}.} \bibinfo{year}{2018}\natexlab{}.
\newblock \showarticletitle{Towards improving CVSS}.
\newblock \bibinfo{journal}{\emph{Software Engineering Institute, Carnegie Mellon University, Tech. Rep}} (\bibinfo{year}{2018}).
\newblock


\bibitem[Spring et~al\mbox{.}(2021b)]%
        {ssvc}
\bibfield{author}{\bibinfo{person}{Jonathan~M Spring}, \bibinfo{person}{Allen Householder}, \bibinfo{person}{Eric Hatleback}, \bibinfo{person}{Art Manion}, \bibinfo{person}{Madison Oliver}, \bibinfo{person}{Vijay Sarvapalli}, \bibinfo{person}{Laurie Tyzenhaus}, {and} \bibinfo{person}{Charles Yarbrough}.} \bibinfo{year}{2021}\natexlab{b}.
\newblock \bibinfo{booktitle}{\emph{Prioritizing Vulnerability Response: A Stakeholder-Specific Vulnerability Categorization (Version 2.0)}}.
\newblock \bibinfo{type}{{T}echnical {R}eport}. \bibinfo{institution}{CARNEGIE-MELLON UNIV PITTSBURGH PA}.
\newblock


\bibitem[Suciu et~al\mbox{.}(2022)]%
        {suciu2022expected}
\bibfield{author}{\bibinfo{person}{Octavian Suciu}, \bibinfo{person}{Connor Nelson}, \bibinfo{person}{Zhuoer Lyu}, \bibinfo{person}{Tiffany Bao}, {and} \bibinfo{person}{Tudor Dumitraș}.} \bibinfo{year}{2022}\natexlab{}.
\newblock \showarticletitle{Expected exploitability: Predicting the development of functional vulnerability exploits}. In \bibinfo{booktitle}{\emph{31st USENIX Security Symposium (USENIX Security 22)}}. \bibinfo{pages}{377--394}.
\newblock


\bibitem[Tai(2020)]%
        {vpr1}
\bibfield{author}{\bibinfo{person}{Wei Tai}.} \bibinfo{year}{2020}\natexlab{}.
\newblock \bibinfo{title}{{What Is VPR and How Is It Different from CVSS?}}
\newblock \bibinfo{howpublished}{\url{https://www.tenable.com/blog/what-is-vpr-and-how-is-it-different-from-cvss}}.
\newblock


\bibitem[Tariq et~al\mbox{.}(2025)]%
        {triagebottlenecks}
\bibfield{author}{\bibinfo{person}{Shahroz Tariq}, \bibinfo{person}{Mohan Baruwal~Chhetri}, \bibinfo{person}{Surya Nepal}, {and} \bibinfo{person}{Cecile Paris}.} \bibinfo{year}{2025}\natexlab{}.
\newblock \showarticletitle{Alert Fatigue in Security Operations Centres: Research Challenges and Opportunities}.
\newblock \bibinfo{journal}{\emph{ACM Comput. Surv.}} \bibinfo{volume}{57}, \bibinfo{number}{9}, Article \bibinfo{articleno}{224} (\bibinfo{date}{April} \bibinfo{year}{2025}), \bibinfo{numpages}{38}~pages.
\newblock
\showISSN{0360-0300}
\href{https://doi.org/10.1145/3723158}{doi:\nolinkurl{10.1145/3723158}}


\bibitem[{Tenable, Inc.}(2023)]%
        {vpr2}
\bibfield{author}{\bibinfo{person}{{Tenable, Inc.}}} \bibinfo{year}{2023}\natexlab{}.
\newblock \bibinfo{booktitle}{\emph{Tenable Security Center 6.1.x User Guide}}.
\newblock \bibinfo{type}{{T}echnical {R}eport}.
\newblock
\urldef\tempurl%
\url{https://docs.tenable.com/security-center/Content/PDF/Tenable_Security_Center-User_Guide.pdf}
\showURL{%
\tempurl}


\bibitem[{The MITRE Corporation}({[n.\,d.]})]%
        {cwe}
\bibfield{author}{\bibinfo{person}{{The MITRE Corporation}}.} \bibinfo{year}{[n.\,d.]}\natexlab{}.
\newblock \bibinfo{title}{{Common Weakness Enumeration}}.
\newblock \bibinfo{howpublished}{\url{https://cwe.mitre.org/}}.
\newblock


\bibitem[{The MITRE Corporation}(2025)]%
        {cve_number}
\bibfield{author}{\bibinfo{person}{{The MITRE Corporation}}.} \bibinfo{year}{2025}\natexlab{}.
\newblock \bibinfo{title}{{CVE Metrics}}.
\newblock \bibinfo{howpublished}{\url{https://www.cve.org/About/Metrics}}.
\newblock
\newblock
\shownote{Accessed: 2025-08-04}.


\bibitem[ThreatGEN({[n.\,d.]})]%
        {ivss}
\bibfield{author}{\bibinfo{person}{ThreatGEN}.} \bibinfo{year}{[n.\,d.]}\natexlab{}.
\newblock \bibinfo{title}{Industrial Vulnerability Scoring System}.
\newblock \bibinfo{howpublished}{\url{https://threatgen.com/resources/ivss/}}.
\newblock


\bibitem[Van~der Maaten and Hinton(2008)]%
        {van2008visualizing}
\bibfield{author}{\bibinfo{person}{Laurens Van~der Maaten} {and} \bibinfo{person}{Geoffrey Hinton}.} \bibinfo{year}{2008}\natexlab{}.
\newblock \showarticletitle{Visualizing data using t-SNE.}
\newblock \bibinfo{journal}{\emph{Journal of machine learning research}} \bibinfo{volume}{9}, \bibinfo{number}{11} (\bibinfo{year}{2008}).
\newblock


\bibitem[Verner et~al\mbox{.}(2009)]%
        {Verner2009}
\bibfield{author}{\bibinfo{person}{J.M. Verner}, \bibinfo{person}{J. Sampson}, \bibinfo{person}{V. Tosic}, \bibinfo{person}{N.A.~Abu Bakar}, {and} \bibinfo{person}{B.A. Kitchenham}.} \bibinfo{year}{2009}\natexlab{}.
\newblock \showarticletitle{Guidelines for industrially-based multiple case studies in software engineering}. In \bibinfo{booktitle}{\emph{2009 Third International Conference on Research Challenges in Information Science}}. \bibinfo{pages}{313--324}.
\newblock
\href{https://doi.org/10.1109/RCIS.2009.5089295}{doi:\nolinkurl{10.1109/RCIS.2009.5089295}}


\bibitem[Vieira et~al\mbox{.}(2010)]%
        {5584447}
\bibfield{author}{\bibinfo{person}{Susana~M. Vieira}, \bibinfo{person}{Uzay Kaymak}, {and} \bibinfo{person}{João M.~C. Sousa}.} \bibinfo{year}{2010}\natexlab{}.
\newblock \showarticletitle{Cohen's kappa coefficient as a performance measure for feature selection}. In \bibinfo{booktitle}{\emph{International Conference on Fuzzy Systems}}. \bibinfo{pages}{1--8}.
\newblock
\href{https://doi.org/10.1109/FUZZY.2010.5584447}{doi:\nolinkurl{10.1109/FUZZY.2010.5584447}}


\bibitem[Walkowski et~al\mbox{.}(2021a)]%
        {walkowski2021automatic}
\bibfield{author}{\bibinfo{person}{Micha{\l} Walkowski}, \bibinfo{person}{Maciej Krakowiak}, \bibinfo{person}{Marcin Jaroszewski}, \bibinfo{person}{Jacek Oko}, {and} \bibinfo{person}{S{\l}awomir Sujecki}.} \bibinfo{year}{2021}\natexlab{a}.
\newblock \showarticletitle{Automatic CVSS-based vulnerability prioritization and response with context information}. In \bibinfo{booktitle}{\emph{2021 International Conference on Software, Telecommunications and Computer Networks (SoftCOM)}}. IEEE, \bibinfo{pages}{1--6}.
\newblock


\bibitem[Walkowski et~al\mbox{.}(2021b)]%
        {walkowski2021vulnerability}
\bibfield{author}{\bibinfo{person}{Micha{\l} Walkowski}, \bibinfo{person}{Jacek Oko}, {and} \bibinfo{person}{S{\l}awomir Sujecki}.} \bibinfo{year}{2021}\natexlab{b}.
\newblock \showarticletitle{Vulnerability management models using a common vulnerability scoring system}.
\newblock \bibinfo{journal}{\emph{Applied Sciences}} \bibinfo{volume}{11}, \bibinfo{number}{18} (\bibinfo{year}{2021}), \bibinfo{pages}{8735}.
\newblock


\end{thebibliography}

\appendix

\section{Correlation Coefficient Analysis}

\begin{figure*}[!thp]
    \centering
\includegraphics[width=1\linewidth]{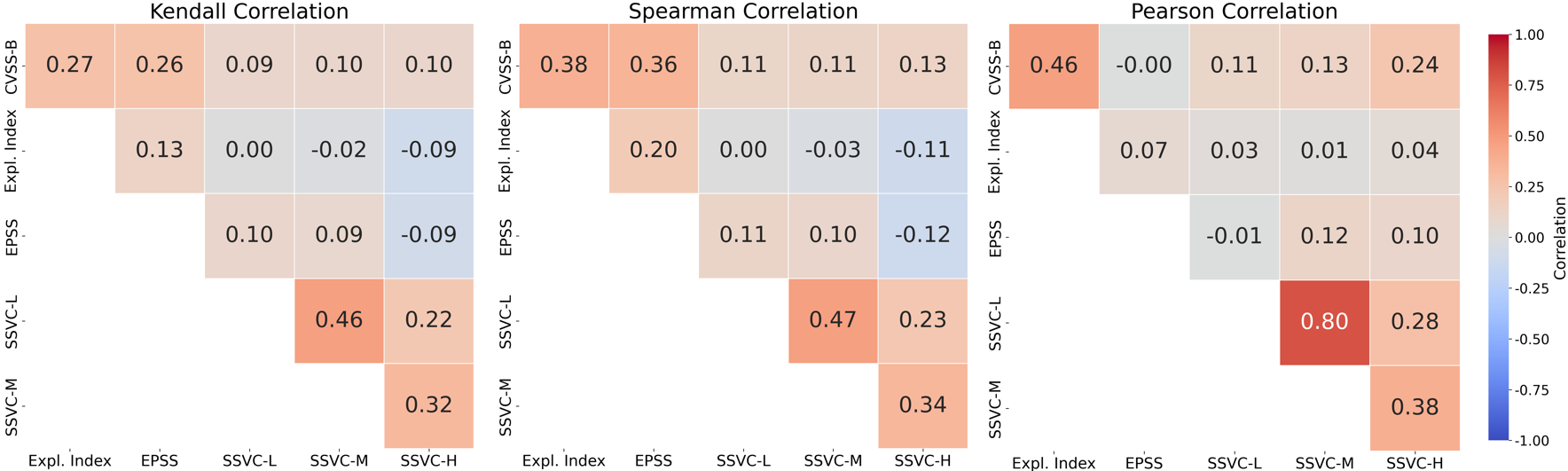}
    \caption{Pearson Correlation, Spearman Correlation, and Kendall's Tau measurements for CVSS, Exploitability Index, EPSS, and SSVC scoring systems.}
    \label{fig:correlations}
    \Description{Pearson Correlation, Spearman Correlation, and Kendall's Tau measurements for CVSS, Exploitability Index, EPSS, and SSVC scoring systems.}
\end{figure*}

Figure~\ref{fig:correlations} demonstrates the Pearson Correlation, Spearman Correlation, and Kendall's Tau measurements for CVSS, Exploitability Index, EPSS, and SSVC scoring systems.

\section{Agreement among scoring systems on top-ranked vulnerabilities}

\begin{figure*}
    \centering
    \includegraphics[width=.9\linewidth]{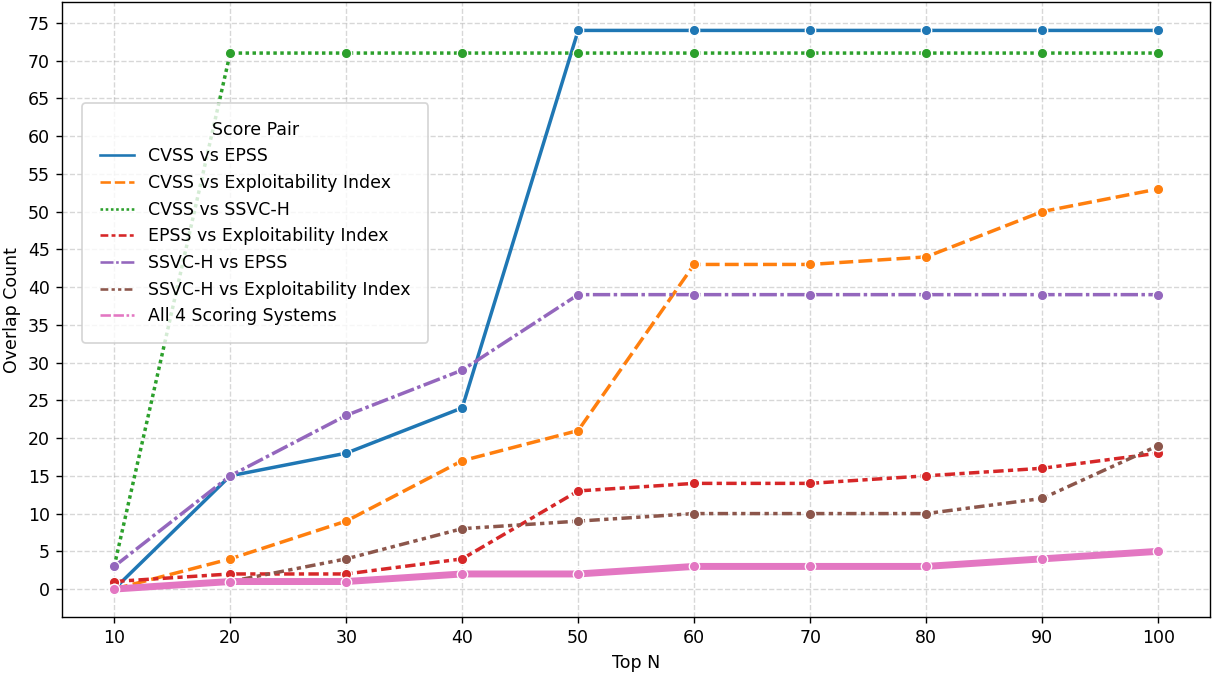}    \vspace{-10pt}

    \caption{Overlap between top-N scored CVEs among scoring system pairs.}
    \label{fig:top-n}

\end{figure*}

Figure~\ref{fig:top-n} demonstrates the results on the agreement among scoring systems on top-ranked vulnerabilities.

\end{document}